\documentclass[lettersize,journal]{IEEEtran}
\usepackage{amsmath,amsfonts}
\usepackage{algorithmic}
\usepackage{multirow}
\usepackage[dvipsnames]{xcolor}
\usepackage{tcolorbox}
\usepackage{booktabs}
\usepackage[linesnumbered,lined,boxed,commentsnumbered]{algorithm2e}
\usepackage{array}
\usepackage[caption=false,font=normalsize,labelfont=sf,textfont=sf]{subfig}
\usepackage{textcomp}
\usepackage{stfloats}
\usepackage{tcolorbox}
\usepackage{url}
\usepackage{framed}
\usepackage{verbatim}
\usepackage{graphicx}
\def\BibTeX{{\rm B\kern-.05em{\sc i\kern-.025em b}\kern-.08em
    T\kern-.1667em\lower.7ex\hbox{E}\kern-.125emX}}
\usepackage{balance}

\usepackage{listings}
\usepackage{xcolor}
\definecolor{commentgreen}{RGB}{2,112,10}
\definecolor{eminence}{RGB}{108,48,130}
\definecolor{weborange}{RGB}{255,165,0}
\definecolor{frenchplum}{RGB}{129,20,83}

\newtheorem{assumption}{Assumption}[section]
\newtheorem{definition}{Definition}[section]
\newtheorem{theorem}{Theorem}[section]

\newtheorem{lemma}[theorem]{Lemma}
\newtheorem{remark}{Remark}

\usepackage{tabularx}
\newcounter{protocol}

\begin{document}

\title{Protecting the `Stop Using My 
Data' Right through Blockchain-assisted Evidence Generation}

\author{
    \IEEEauthorblockN{Fan Zhang\IEEEauthorrefmark{1}, Peng Liu\IEEEauthorrefmark{1}}\\
    % \IEEEauthorblockA{\IEEEauthorrefmark{1}ByteDance Inc.
    % \\\{1, 4\}@abc.com}
    \IEEEauthorblockA{\IEEEauthorrefmark{1}Pennsylvania State University}
    %\\\{pxl20\}@psu.edu}

\IEEEcompsocitemizethanks{\IEEEcompsocthanksitem Fan Zhang and Peng Liu are with Penn State University, State College,
PA, 16801 (email: fxz5095@psu.edu, pxl20@psu.edu).}}
\maketitle

\lstset{
    % language=Basic,
    frame=tb,
    tabsize=4,
    breaklines      =   true,
    columns         =   fixed,
    showstringspaces=false,
    % numbers=False,
    %upquote=true,
    commentstyle=\color{commentgreen},
    keywordstyle=\color{eminence},
    stringstyle=\color{red},
    basicstyle=\small\ttfamily, % basic font setting
    emph={int,char,double,float,unsigned,void,bool},
    emphstyle={\color{blue}},
    escapechar=\&,
    % keyword highlighting
    classoffset=1, % starting new class
    otherkeywords={>,<,.,;,-,!,=,~},
    morekeywords={>,<,.,;,-,!,=,~},
    keywordstyle=\color{weborange},
    classoffset=0
}

\section*{Abstract}

In order to provide personalized services to users, 
Internet-based platforms collect and utilize user-generated behavioral 
data. 
Although the `stop using my data' right should 
be a fundamental data right, which allows 
individuals to request their personal data to be no longer 
utilized by online platforms, 
the existing preventive data protection measures 
(e.g., cryptographic data elimination, differential privacy)  
are unfortunately not applicable.  
This work aims to develop the first Evidence Generation
Framework for deterring post-acquisition data right violations. 
We formulated the `stop using my
data’ problem, which captures a vantage facet of
the multi-faceted notion of `right to be forgotten’. 
We designed and implemented the first blockchain-assisted
system to generate evidence for deterring the 
violations of the ‘stop using my data’ right. Our system
employs a novel two-stage evidence generation protocol
whose efficacy is ensured by a newly proposed Lemma. 
To validate our framework, 
we conducted a case study on recommendation
systems with systematic evaluation experiments
using two real-world datasets: the measured
success rate exceeds 99\%. 

%This paper introduces the first blockchain-assisted evidence generation system to protect the `stop using my data' right of users. In scenarios where a user's data, despite requests to cease its use, continues to be exploited by a platform, our developed blockchain-based system creates evidence by analyzing the influence of user data on platform outputs, ensuring the cessation of specified data use. We selected recommendation systems as our case study and carried out data analysis from well-known platforms such as Amazon. We proposed varying levels of protection schemes for different users. Our experimental results demonstrate that our approach effectively monitors the system and upholds users' `stop using my data' rights with a success rate exceeding 99\% for target users. 
\section{Introduction}
\label{sec:intro}

The advent of Internet-based platforms and applications has intensified the prevailing concerns about users' {\bf rights} on personal data, thereby posing an enduring research challenge. Despite the research community's persistent efforts, which have resulted in a plethora of system and protocol design proposals, online platforms and applications continue to utilize users' behavioral data (e.g., purchase activities), leaving important data right protection issues unaddressed 
% avoiding detection and traceability 
\cite{sha2020survey, mayrhofer2021android}. 
As reported in article \cite{Termly2024},  
63\% of Internet users believe most companies are not transparent 
about how their data is used. 
%data indicates that more than 30\% of individuals are worried about the improper use of their personal data. 
For instance, consider a scenario where you request an e-commerce platform to stop using your personal shopping records. If this data right protection 
request is taken seriously, the recommendation system running 
on the platform would cease to 
suggest products/items derived from your prior purchases. 
Contrarily, the statistics obtained by various studies, including 
\cite{Termly2024}, reveal a significant deviation. 
Despite submitting the request, you may notice that the 
recommendation system continues to recommend items linked with 
your past preferences \cite{rocha2020functionality}. 
Such a situation not only violates users' control over their personal data 
but also poses significant threats to the sanctity of data rights 
placed by law (and regulations) \cite{jung2020secure,li2022design}. 
%This deficiency is largely due to a lack of robust security measures and awareness. 

% The `stop using my data' rights,allows users to request that their personal data be erased from the recommendation system database. This provision enables users to have control over their personal data, including the right to withdraw their consent anytime. Unfortunately, data leakage is still an alarming issue in recommendation systems. It can happen through cyber-attacks or human mistakes, resulting in the unintentional disclosure of personal data to unauthorized parties. The ramifications to users are immense as it can lead to identity theft, fraud, and other malicious activities. 

In order to deal with the potential data right violations, 
several frameworks have been proposed to protect user data rights after data has been collected, embedding concepts such as `the right to be forgotten',  
which empowers users to demand the deletion or removal of personal data from online platforms, especially when the data in question is obsolete, irrelevant, or no longer serves its original purpose \cite{zaeem2020effect}. 
The European Union has established this right 
through legal frameworks such as the General Data 
Protection Regulation (GDPR) \cite{voigt2017eu, bygrave2019minding}. 
However, it turns out to be very challenging to guarantee 
 that every `stop using my data' request is indeed honored 
 by the corresponding online platform. 
%A primary root cause of this daunting challenge is that 

%This complexity implies that enforcing ``the right to be forgotten'' represents a considerable challenge, primarily because conclusive verification of total data deletion remains a technically convoluted task. 

Regarding why it is very challenging to guarantee, 
although one main reason is that most companies are not transparent 
about how their data is used, lack of sufficient  
data right protection technologies is obviously another main reason. 
Previous technological works 
have primarily concentrated on strategies to 
eliminate users' personal data \textbf{prior} to its utilization. 
This has been predominantly achieved through the use of 
encryption techniques and the principle of differential 
privacy \cite{ lecuyer2019certified, dong2022gaussian}. 
Data encryption, a topic diligently studied over the years, consists 
of cryptographic systems and protocols. 
%Through various mathematical models and transformations, 
These systems and protocols ensure data elimination  
% confidentiality and integrity of data 
by {\em preventively} converting data into an useless format: only the parties 
knowing the decryption key(s) can make the data useful again   
%that can only be deciphered using a decryption key 
\cite{alabdulatif2020towards, das2021bim}. 
Differential privacy (e.g., \cite{wang2022srr,wang2023shuffle}), 
on the other hand, is another {\em preventive} measure 
that primarily deals with adding a certain amount of random noise to the data. 
Although differential privacy mechanisms do not directly 
achieve data elimination, they enable users to hide 
Personal Identifiable Information (PII) from online platforms. 
%This essentially masks the identity or characteristics of individual data entries while maintaining the overall quality and usability of the dataset for analytical tasks \cite{wang2022srr, wang2023shuffle}. 

Despite these remarkable advances, a glaring {\bf gap} in the 
literature pertains to the lack of strategies addressing the 
data right protection needs arising from post-acquisition data utilization. 
Because the user behavioral data is being routinely acquired 
by online platforms, all the aforementioned preventive 
technologies, including differential privacy, would become 
{\bf not applicable}. 
In essence, while preventative mechanisms are progressing, technologies 
%corrective and control measures 
addressing the data right protection needs 
once data has been obtained and utilized 
are still largely unexplored in the literature. 

In this paper, we seek to develop the first Evidence Generation 
Framework for deterring post-acquisition data right violations. 
In particular, we formulate the {\bf ‘stop using my data’ problem} 
by advocating that this right is a fundamental aspect of data protection,
allowing individual users to request their behavioral data
to be no longer utilized by a particular platform. 
Note that since this problem is restricted to users' right on 
whether behavioral data (e.g., what items are purchased and when) 
can be utilized by online platforms, our 
problem is orthogonal to PII protection, and therefore, our frmework  
is {\bf not} solving a privacy problem. 
%it is {\bf NOT} a privacy-enhancing technology (PET). 

%we defined `stop using my data right', focusing on the usage of data and implementing technological oversight of data usage. We aim to empower the individual to halt the consumption of their personal data, thereby creating a more controlled digital environment. We are not trying to maintain user privacy from the perspective of isolating data acquisition, but to protect data rights from the perspective of data usage. We believe that data rights not only include privacy rights but also usage rights. \textbf{Our study diverges from existing research in the following ways}:

 %(a) Our research does not primarily focus on the concept of `the right to be forgotten' due to the intrinsic complexities associated with verifying the cessation of the system's usage of the data. 
 %Alternatively, we introduce the principle of 'stop using my data right', the validity of which can be ascertained by evaluating whether the system persists in utilizing the subject data through systematic output analysis.\cite{kuperberg2020towards, mangini2020empirical}. 

Since preventive technologies, including differential privacy, 
are not applicable, our framework focuses on reactively 
deterring data right violations. 
In particular, in our threat model, 
after receiving the ``stop use my data''
request from a user, if the recipient platform lets its algorithm 
(e.g., recommendation system) 
continue utilizing the user’s behavioral data, the platform is
an attacker. Accordingly, our goal is as follows: through an evidence generation
system, the evidence generator aims to collect compelling
evidences showing that the suspect is indeed an attacker. 

{\bf Technical challenges. } 
{\em Data Accessibility:} As mentioned earlier, most companies are not 
transparent about how their data is used. As a result, 
%It is clear that if the attacker platform  always generates identical outputs since the request was issued, the real workflow must be violating the user’s data right. 
%However, this straightforward criteria cannot be utilized due to poor feasibility. Because the platform is violating the user’s data right, we must not assume that the evidence generation system is endorsed by the platform. 
%the outputs of neither honest platforms, who will honor every ``stop using my data'' request, not attacker platforms, who will continue utilizing users' data, are known to the evidence generation system. 
the platform inherently has
accessibility to all user behavioral data. In contrast, potential
victim users can only access their personal data, remaining
oblivious to the extensive data owned by others. This disparity
introduces a major technical barrier for the evidence generation system.

{\em Grey-box System: } Each platform is a grey-box:
the outputs can be observed, but the platform is not
transparent about how the data is used. 
%is impossible to track what behavioral data is utilized (by the algorithm) and what is not. When certain metadata derived from the behavioral data is utilized by the algorithm, the metadata is also kept confidential from outsiders.

{\em The Lie of User: } Users could forge information to defraud
the platform compensation, for example, by claiming 
data right violation without actually asking the platform to stop
using their data.

{\bf Approach overview. }
To address the challenge of data accessibility, our framework
proposes the creation of special Web3 communities.  
Since every user who regularly uses the 
platform could become a potential victim of the violation of
his or her `stop using my data’ right, one or more Web3 communities 
can enable the potential victims to collectively obtain improved data
right protection. Each such Web3 community corresponds 
to one subnet of users. 
%with similar user behavior (e.g., the users may buy merchants belonging to the same category).
Each Web3 community leverages the blockchain technology to 
maintain a shared immutable record of user behavioral data, and 
to foster a decentralized trust-based community for the potential victims. 
To ensure privacy, the blockchain nodes do not hold any 
  cleartext data; rather, hash values of user data are stored. 
%needs to utilize specific user behavioral data while making sure that the user privacy is preserved. For this purpose, our framework adopts a potent resolution - the establishment of a Multi-Party Computation (MPC) group. 

To address the challenge of grey-box systems, our framework
leverages the approximations (of the black-box components)
achieved through the user behavioral data jointly collected by 
each Web3 community and the computation results of  
secure Multi-Party Computation (MPC). Note that MPC is 
adopted to enable each Web3 community's evidence generation system to 
collectively generate approximate platform outputs in
a privacy-preserving manner. 
Finally, to address the challenge of lie of user, our framework
employs blockchains to maintain an immutable record of
user behavior and the platform’s outputs, thus offering
transparency and mitigating the potential dishonest practices.

%Reflecting on the issue previously mentioned, our first step was to clearly define and restrict the platform. We maintained the stance that only those systems that rely on user input to generate outputs can be scrutinised. Following this, we created a system that generates evidence, which was based on both user input and system output. This system's goal was to oversee and collect proof of any incorrect use of user data. To better understand how system output and user input are linked, we assembled a web 3 community. The data provided by this community helped us estimate the functions of the system. In summary, our {\bf main contributions} are as following:
% 1) We conducted a comprehensive analysis of data from various websites to determine whether personal information is being erased. The results of our analysis are presented in Table 1. 2) Our research is the first to offers the systematic modeling analysis of this particular problem and introduce a groundbreaking protection detection system. 3) We conducted experiments on diverse datasets and found that our system yields a high detection accuracy, achieving a performance level of xxx.

In summary, our main contributions are as follows: 
\begin{itemize} 
    \item This work is the first that formulates the `stop using my data' problem, which captures a vantage facet/aspect of the multi-faceted notion of `right to be forgotten'.  
    \item We proposed the first evidence generation framework for 
    solving the `stop using my data' problem. 
    \item We designed and implemented the first blockchain-assisted system to generate evidence for deterring the violations of the `stop using my data' right. The system employs a novel two-stage evidence generation protocol whose 
    efficacy is ensured by a newly proposed Lemma. The proof is also provided.   
    \item Towards validating the proposed evidence generation framework, we conduct a case study on recommendation systems with systematic evaluation experiments using two real-world datasets (i.e., the Amazon Magazine dataset and the Amazon Beauty dataset). The measured success rate exceeds 99\%. 
    %\item We integrated the Web3 community into our system design, thus ensuring the independence and trustworthiness of the user group. Moreover, we obtain information via the MPC system without disclosing personal user data.
    %\item Our selection of recommendation systems underwent validation through the use of real-world datasets from Amazon. The effectiveness of our system is demonstrated by its capability to detect and monitor data misuse, achieving a success rate exceeding 99\%. 
\end{itemize} 

%https://www.statista.com/statistics/1382859/concerns-personal-data-misuse-by-age-and-gender/
\section{Background}
\label{sec:back}

\subsection{The Right to Be Forgotten} 
The right to be forgotten is a legal concept that allows individuals to request the removal of their personal information from public records and 
Internet-based platforms (e.g., E-commerce platforms, social media platforms). 
% It is also known as the right to erasure. 
This right was introduced by the 
European Union's General Data Protection Regulation (GDPR) in 2018 \cite{politou2018forgetting, peloquin2020disruptive}. The regulation gives individuals the 
right to have their personal data erased under certain 
circumstances, such as when the data is no longer necessary for 
the purpose it was collected, or when a person withdraws her 
consent for the data to be processed \cite{zaeem2020effect}. 

To exercise the right to be forgotten, individuals must make a formal request to the organization storing their personal data \cite{politou2018forgetting}. The organization must then assess whether the request is valid and take appropriate actions, which may include deleting the data or making it anonymous. Implementing the right to be forgotten can be challenging, particularly in the digital age where data can be easily copied and disseminated. Some solutions include data anonymization, encryption and the use of technology to monitor and remove personal information from the Web. 

%------------------------
\subsection{Blockchains}

The Blockchain technology is revolutionizing the way in which decentralized 
systems are built. 
%transactions are recorded and verified. 
%It is a decentralized, distributed database that eliminates the need for intermediaries, providing a secure and transparent system for conducting transactions \cite{sanka2021systematic, monrat2019survey}. 
A blockchain is a distributed ledger that records transactions 
in a transparent, immutable way. Through immutable transaction 
recording, various communities of stake holders (e.g., 
buyers and sellers of Bitcoins) can track not only 
tangible assets such as cash and land but also 
intangible assets such as patents, copyrights, and branding. 
A most notable property of the Blockchain technology 
is that trustworthiness is achieved without 
the need for intermediaries (e.g., trusted third parties)
\cite{huynh2023blockchain, bhutta2021survey}. 
The ledger consists of a chain of blocks. Each block contains a unique 
hash value that identifies itself and is linked to its previous block. 
% creating an unbreakable chain of information .

An essential feature of the blockchain technology is its operations on a peer-to-peer network, achieving decentralized data management \cite{li2020survey}. This implies that rather than a singular entity owning and controlling all data, the data is disseminated across numerous nodes in the network.
The decentralization feature is critical for dealing with 
 cyber threats. Any modification attempt against a block's content necessitates a consensus from the majority of the network - a task that is practically and computationally infeasible \cite{choi2022blockchain}. 
Since blockchains provide a highly desirable way to track 
all kinds of assets, 
% These unique features of blockchain technology extend its application beyond the common financial transactions in cryptocurrency. 
the technology has penetrated into various industry sectors, including finance, supply chain management and healthcare.
%demonstrating its potential to revolutionize numerous sectors. 

%------------------------------------
\subsection{Multi-party Computation}
Multi-party computation (MPC) is a cryptographic technique that allows multiple parties to jointly compute a function on their private inputs without revealing any information about them \cite{damgaard2006scalable, alexopoulos2017mcmix}. This technique is useful in scenarios where the parties do not trust each other, but need to collaborate to achieve a common goal. 
The concept of MPC was first introduced in the 1980s, with the development of secure function evaluation protocols by Andrew Yao \cite{yao1982protocols}. Some of the most widely used MPC protocols include the Secure Multi-Party Computation Protocol (SMC), the Yao’s Millionaires Problem Protocol, and the Garbled Circuit Protocol. 

MPC involves several steps \cite{zhao2019secure}. First, the 
parties agree on a protocol to use for computation. 
%This protocol usually involves a series of messages exchanged between the parties, with each message containing a partial computation of the function. 
Second, each party privately inputs their data into the protocol. 
Third, the parties exchange messages according to the agreed 
protocol to compute the desired function. 
%Finally, each party receives the output of the computation without knowing the inputs of the other parties. 
% To achieve secure computation in MPC, different cryptographic techniques are used \cite{zheng2020privacy}. For example, secret sharing can be used to split a secret among the parties such that the secret can only be reconstructed when a threshold number of parties collaborate. Secure function evaluation (SFE) can also be used to allow parties to compute a function securely without revealing their inputs. Homomorphic encryption is another technique that can be used to perform computations on encrypted data without the need for decryption. 
MPC techniques have been widely used in various domains, such as secure 
data analysis, privacy-preserving machine learning, and secure multiparty computation. 
%The model has proven to be effective in protecting the privacy of participants and maintaining the confidentiality of data. 

%----------------------------
\subsection{Web3 Communities}
With increasing concerns on data privacy, security, and control, the Web3 movement signifies a shift in paradigm towards a web that is more decentralized and user-centric \cite{fan2023altruistic}. Web3 builds upon the previous iterations of the Internet, namely Web1 and Web2, to address the limitations of these centralized models \cite{nabben2023web3}. By leveraging the innovative capabilities of the blockchain technology, Web3 aims to empower individuals, granting them sovereignty over their own digital identities, data, and transactions. 

Distributed ledgers lie at the core of Web3, enabling the creation of decentralized systems that facilitate secure and transparent transactions \cite{farahani2021convergence, zhu2019applications}. Smart contracts, self-executing pieces of code, further enhance the capabilities of Web3 by automating and enforcing agreements without the need for intermediaries \cite{zou2019smart}.
% The Web3 movement has been resulting in a variety of decentralized applications (DApps) and Web3 Communities.
%play a crucial role in driving the development and adoption of Web3 technologies. 
%Through collaboration, knowledge sharing, and open-source contributions, a vibrant Web3 community fosters innovation and explores the possibilities offered by decentralized applications (DApps), protocols, and governance systems. 
A Web3 community is a virtual community where a group of people with shared 
digital ownership can work together on achieving common goals. 
\section{The ``Stop Using My Data'' Problem}

\label{sec: problem}
In this section, we formulate the `stop using my data' problem. The problem is defined based on a general platform model. In addition, we introduce the threat model to further explain the capabilities of attackers and defenders. After that, we design a third-party judge model. Lastly, the challenges inherent in solving the newly formulated problem are discussed. 

\subsection{Platform Model}
Our platform model intends to cover a variety of 
internet-based platforms that   
%Platforms refer to e-commerce websites or applications that 
provide users (e.g., anyone who is at least 18 years old can use Amazon) with meaningful services.
% or facilitate transactions 
Such platforms may include marketplaces, social media platforms, content sharing websites, online service providers, and so on. 

Since this work focuses on protecting private 
{\em behavioral} data instead of PII (personally identifiable information), 
%To protect the data of users through information of the output, 
the platforms should have the following 
properties: 1) The operations of the platforms involve not only 
certain behavioral data (of every user) but also an algorithm 
that uses the behavioral data as a main input for generating 
the output shown to a user. 
For instance, online marketplaces generate recommendations based on the users' purchase histories.
%and the recommendations are real-world items. 
For another example, 
the Google keyboard system makes query suggestions based on a user's previous input habits. 
%We consider such platform as our target platforms. 
%2) The output of the platforms are determined by the input, and the output or segmented output belongs to a finite set, which is publicly available. 
2) The algorithm is a white-box and known to all users, 
but the decisions on whether to utilize a particular user's behavioral data 
are made in a black-box (i.e., unknown to users).  
%the algorithm and the data, among which the algorithm is a white-box model, and the data is a black-box model.  

Based on these properties of a platform, we give the definition of the `stop using my data' right:

\begin{definition} [Stop Using My Data Right] 
The `stop using my data' right is a fundamental aspect of data protection, allowing individual users to request that their  behavioral data no longer be utilized by a particular platform.
\end{definition}

% \textcolor{red}{add one paragraph about attack, begin with bold font}

% \begin{definition} [Stop Using My Data Right] 
% The `stop using my data' right is a fundamental aspect of data protection, allowing individuals to request that their personal data no longer be utilized by a platform.
% \end{definition}

% \textcolor{red}{need another more general model: the platform model, so people know what is the platform (section 4.1), inputs, outputs, data, algorithm(system)--the term used in Google keyboard;
% 1. identify the definition; only using the terms the authors have already introduced, a definition could be ;(rigorious)}

% \subsection{Attacker Model}

\subsection{Threat Model}
The definition of the `stop using my data' right indicates 
that the data right protection problem may involve the following parties: 

%We consider a user-platform model with two parties: The target platform who collects information from a user subsequently tailors its services based on this data. The user who requests that the platform cease the use of his personal information but hard to judge whether the platform continue using the data. We give the definitions of both attacker and protector in the following. 

\textbf{Attacker \& Victim}: After receiving the ``stop use my data'' request from a user, if the recipient platform lets its 
algorithm continue utilizing the user's behavioral data, the platform is an attacker. Accordingly, the user is the victim. 

% Defined as the target platform, the attacker continues to exploit user information despite requests to desist.

\textbf{Evidence Generator}: 
Through an evidence generation system, the evidence generator 
aims to collect compelling evidences showing that the 
suspect is indeed an attacker. Note that the victim 
is usually only part of the entire evidence 
generation system. 
%This entity, integral to an evidence generation system, aims to ascertain whether a platform is acting maliciously by gathering evidence to verify any violation of the users' `stop using my data' right.

\textbf{Judge}:  
The judge is supposed to examine the collected evidences
and decide if the accused is guilty.

In this threat model, the attacker possesses access to all 
of the behavioral data used in running the 
platform's algorithm, whereas the evidence generation system 
% defender 
is usually only utilizing a small subset of the behavioral data.  
% specified users' data. 
For example, when recommendation systems are using 
user behavioral data to make recommendations, 
%running a collaborative filtering algorithm, 
the inaccessibility of the database of user purchase 
histories prevents outsiders (e.g., the victim 
and his or her allies) 
from accurately determining whether 
the recommendations are made 
based on the victim's purchase history or not. 

% This study comprises two distinct models: a `Grey-Box' model, in which the recommendation algorithm is public, and a `Black-Box' model, in which the entire system is unknown to the protector. Based on these paradigms, we apply various methodologies and experimental techniques. 

%------------------------
\subsection{Judge Model}

The judge, which is potentially a court of law, 
is supposed to examine the collected evidences
  and decide if the accused is guilty.
Since the attacker is not assumed a honest party, the accused platform 
 may certainly deny any ``violation of data right'' accusations.  
Moreover, the users who raise an accusation could also be lying. 

Therefore, the Judge Model should take 
all possible consequences of dishonesty into consideration. 
%We developed a mechanism to assess the adequacy of evidence produced by an evidence-generation system. This mechanism, namely judge model, potentially a court of law or another impartial entity, faces challenges related to the accuracy of information provided by the system or its users. 
In particular, the Judge Model assumes that in order to obtain more benefits (e.g., profits), a platform could violate user data rights when processing their behavioral data. Alternatively, users might baselessly claim their data has been misused to pursue monetary 
compensations. Thus, the Judge should focus on two major tasks: 1) Verifying if the platform has ceased using user data based on the available evidence; 2) Identifying if a user's allegations of unauthorized data use for compensation are unfounded. 
To fulfill the first task, the Judge relies on the evidences from the evidence-generation system. %to confirm compliance with regulations. 
To fulfill the second task, it is desirable to maintain a decentralized 
yet trustworthy record of each user's behavioral data. 
Because it is decentralized, the Judge does not need to trust any 
third parties. Because it is trustworthy, the Judge can assume that users 
cannot fabricate (or destroy) any evidences.  
Note that if the trustworthiness comes from the consensus of a peer-to-peer network, 
then the corresponding privacy-preserving requirements must be met. 

%maintaining a record of each user's requests is crucial. We addressed this challenge by integrating a blockchain system that thwarts attempts by users to fabricate evidence. 
%To obviate the need for reliance on any centralized database, the implementation of a blockchain system is necessary. Only in this system can we monitor the flow of information while ensuring the independent storage of user data.

%--------------------------
\subsection{Challenges}

\begin{figure}[htbp]
    \centering  \includegraphics[width=8cm]{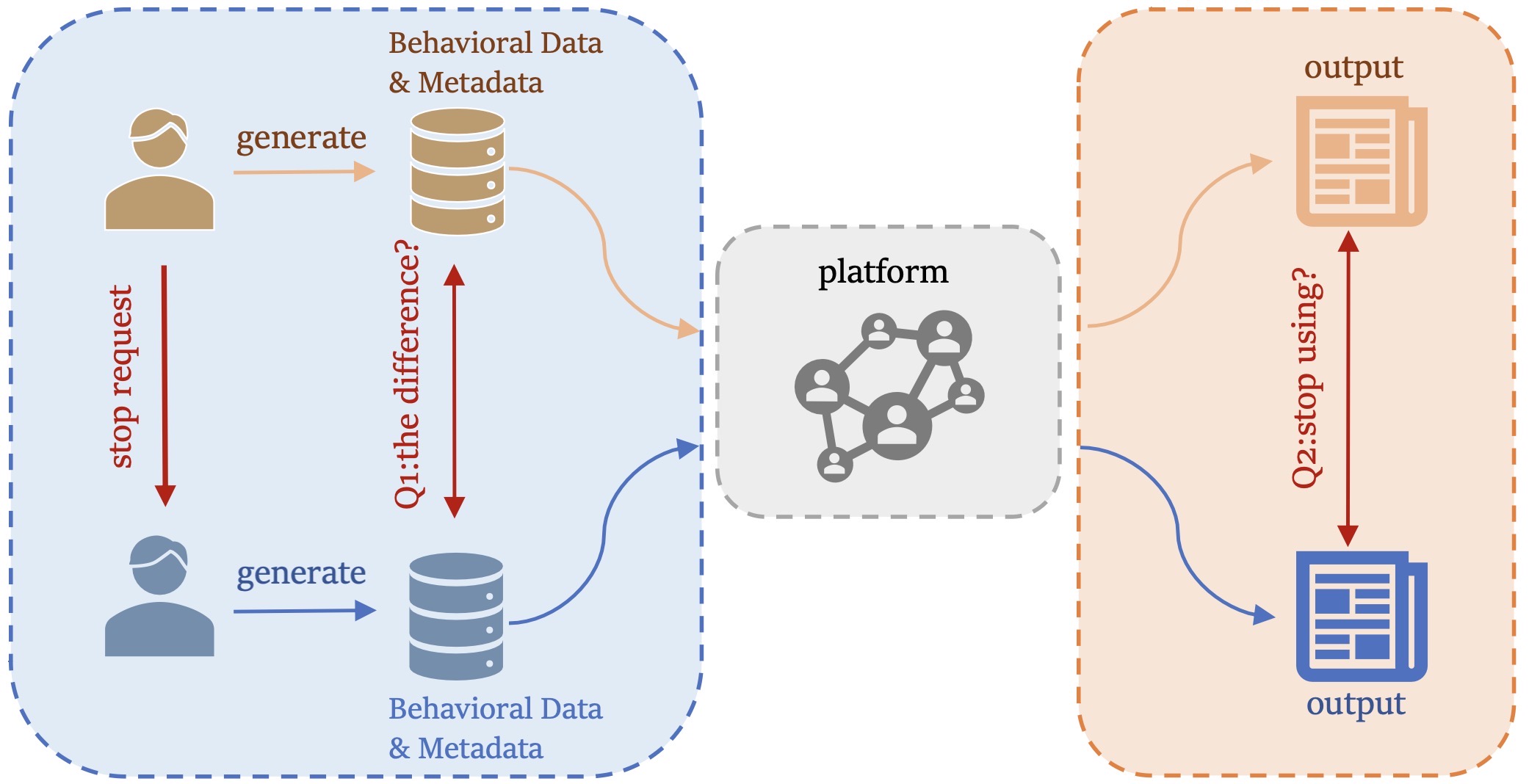}
    \caption{The problem.}
    \label{fig:problem}
    \vspace{-0.1in}
\end{figure}

A simplified illustration of the `stop using my data'  problem is 
shown in Figure \ref{fig:problem}, where  
% illustrating that it principally centers on two distinct issues.
only one user has issued the `stop using my data' request. 
Accordingly, after the request was issued, 
  the brown color components are part of the worst case workflow 
  (of the platform) that totally ignores the request.  
In contrast, the blue color components are part of the 
  workflow that is really implemented by the platform. 

It is clear that if the real workflow and the worst case workflow 
 always generate identical outputs since the request was issued, 
 the real workflow must be violating the user's data right. 
However, this straightforward criteria cannot be utilized due to 
poor feasibility. Because the platform is violating the 
user's data right, we must {\em not} assume that the 
evidence generation system is endorsed by the platform. 
Accordingly, the outputs of the worst case workflow 
are never known to the evidence generation system. 

%\begin{tcolorbox}[title=\textbf{Two-phase Problems}, colback=red!5,colframe=red!75!black]
%Q1. How to choose the post-STOP-request probing item? 

%Q2. How to judge whether the platform stops using the data?
%\end{tcolorbox}

Inability to observe the worst case workflow's outputs is actually only 
one of the following challenges we face in solving the 
`stop using my data' problem. 
%To solve these questions, there are several challenges we need to consider:
% sixth paragraph
\label{sec:challenge}

{\bf Data Accessibility \label{challenge: data}}: The disparity in accessible data represents a significant challenge. The platform inherently has accessibility to all user behavioral data. 
In contrast, potential victim users can only access their personal data, remaining oblivious to the extensive data owned by others. 
This disparity introduces a major technical barrier for the 
evidence generation system. 
%complicates the detection process. 

{\bf Grey-box System}: 
\label{challenge: system}
%Grey-box systems refer to systems where the algorithm and theory is public but parameters cannot be accessed, making it impossible to track how data is utilized. 
Each platform is a grey-box system: the algorithm is public, the outputs can be observed, but it is impossible (for outsiders) to track what behavioral data is utilized (by the algorithm) and what is not. When certain metadata derived from the behavioral data is utilized by the algorithm, the metadata is also kept confidential from outsiders. 
%It is essential to prioritize protecting users' right to require stopping using their data because of the inability to determine the storage status of users' data, which means that their `right to be forgotten' cannot be guaranteed. 
% In this scenario, even if we only focus on the use of data, the challenge of measuring the correlation between system output and previous data usage remains daunting due to the grey-box internal function of the system.

{\bf The Lie of User}: 
\label{challenge: user}
Users could forge information to defraud the platform compensation, for example, by providing false requests (when in reality they did not ask the platform to stop using their data). In this context, we need to ensure that the user is honest and provides truthful data. 

\section{Framework Design}

\subsection{Overview}

In this section, we present a framework to address the 
data right protection 
challenges identified in the previous section. 
%A main aspect of the the framework is the functions and design principles for each component.  
To address the challenge of data accessibility, our framework
lets a group of users to collaboratively collect their 
behavioral data through the creation of a special Web3 community. 
%proposes the creation of special Web3 communities.  Since every user who regularly uses the platform could become a potential victim of the violation of his or her `stop using my data’ right, one or more Web3 communities  can enable the potential victims to collectively obtain improved data right protection. Each such Web3 community corresponds to one subnet of users. 
Each Web3 community leverages the blockchain technology to 
maintain a shared immutable record of user behavioral data, and 
to foster a decentralized trust-based community for the potential victims. 
%To ensure privacy, the blockchain nodes do not hold any cleartext data; rather, hash values of user data are stored. 
To address the challenge of grey-box systems, our framework
leverages the approximations (of the platform's black-box components)
achieved through  secure Multi-Party Computation (MPC). MPC is 
adopted to enable each Web3 community's evidence generation system to 
collectively generate approximate platform outputs in
a privacy-preserving manner. 
Finally, to tackle lie of user, our framework
employs blockchains to maintain a shared immutable record of
user behavior and the platform’s outputs, thus offering
transparency and mitigating the potential dishonest practices.

%To address the challenge of \textbf{data accessibility}, the framework needs to utilize specific user behavioral data while making sure that the user privacy is preserved.  
%For this purpose, our framework adopts a potent resolution - the establishment of a Multi-Party Computation (MPC) group.  To address the challenge of \textbf{grey-box systems}, our framework leverages the approximations (of the black-box components) achieved through the user behavioral data utilized by the framework and the computation results of the MPC group. 
%Finally, to address the challenge of \textbf{lie of user}, our framework employs blockchains to maintain a shared immutable  record of user behavioral data and the platform's outputs, thus offering transparency and mitigating the potential dishonest practices. To ensure privacy, the blockchain nodes do not hold any cleartext data; rather, hash values of user data are stored in the blockchains. 

\begin{figure}[t]
    \centering  \includegraphics[width=8cm]{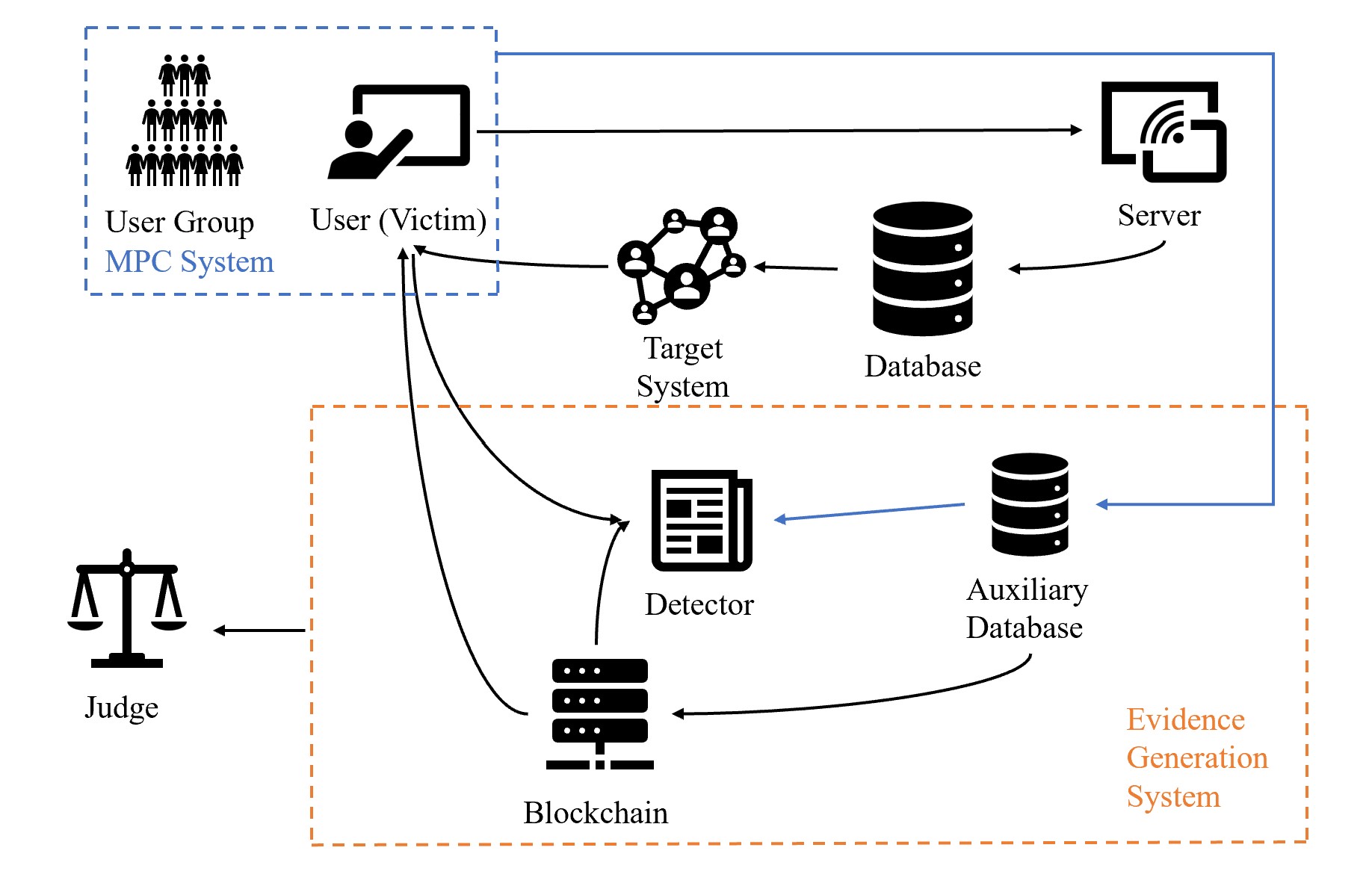}
    \caption{The evidence generation framework.}
    \label{fig:sys}
    \vspace{-0.1in}
\end{figure}

%Given the myriad challenges described earlier, simplifying the system proves to be difficult. 

The main components of our framework are shown in Figure \ref{fig:sys}:  
%presents the complete system, comprising nine elements - the victim user, the MPC Group, server, target system, dataset, auxiliary dataset, blockchain, detector, and the judge. In the following section, we will provide a brief overview of each part: 

\textbf{User Group}: Since every user who regularly uses the platform 
could become a potential victim of the violation of his or her `stop using my data'
right, our framework facilitates the creation of 
one or more special Web3 communities for the potential victims 
to {\em collectively} obtain improved data right protection. 
Each such Web3 community corresponds to one group of users 
(of the platform) with similar user behavior (e.g., 
the users may buy merchants belonging to the same category). 
Since our framework employs 
blockchains to maintain a shared immutable  
record of user behavioral data, Web3 communities are 
a natural way to foster a decentralized 
trust-based community for the potential victims.  
%This group forms part of , contributing data to an MPC system and fostering a trust-based community built on blockchain technology.

\textbf{The Victim User}: The victim user is a member of one of the 
aforementioned Web3 communities. The victim user has asked 
the platform to stop using his or her user behavioral data, 
but unfortunately his or her request has been ignored. 
%Users interact with the platform, generating data, including sensitive information, during this process. Consequently, they may request the cessation of certain data usage to protect their privacy. 

\textbf{Server}: The server is a main component of the platform.  
%feedback, storing this information in a secure database. 
It establishes the network connections with each user side device. 
%It provides GUI to interact with users. 
It forwards the outputs of the Target System to users. 
It also collects the relevant user behavioral data.  

\textbf{Database}: The collected user behavioral data are stored 
in the database. In addition, certain metadata (e.g., the similarity map 
used in recommendation systems) is also stored 
in the database. 
%It stores all data produced by users, facilitating its transfer to the intended systems. 

\textbf{Target System}: This system is another main 
component of the platform. If the platform decides to utilize 
user behavioral data, the target system will run 
particular algorithms (e.g., an algorithm used to 
generate recommendations) to 
% This system accesses user data from the database 
% to deliver personalized information to the users based on their records. 
% It provides the services the platform intends to provide, 
process user behavioral data for making the services the 
platform provides   
more personalized and more engaging.
Note that the outputs of the target system 
are delivered to users through the aforementioned Server. 

\textbf{Auxiliary Database}: Being part of the Evidence 
Generation System, the auxiliary database is {\em not} 
part of the platform. It comprises the data 
or metadata utilized by the Detector. 
The database is built through the Web3 community 
the victim user resorts to.  
%including historical data or publicly available datasets from compromised users.

\textbf{Detector}: When an user thinks that he or she 
has become a victim, the Detector utilizes the Auxiliary Database 
to probe the platform and detect whether the user's `stop using my data' right 
has been violated. Both the detection logic and the 
observations obtained by the detector are major evidences 
the Evidence Generation System will later present to the Judge. 
In addition, since the victim user is directly benefited 
by the Detector, we assume that the victim user has 
strong incentives to run the (lightweight) Detector software on his or her own device. 
However, it should be noticed that in case the user is 
dishonest, he or she could manipulate the Detector software. 
To address this kind of dishonesty, our framework provides the Judge with 
the capability of re-executing the Director software.  

%Our threat model assumes that the Detector is trusted by the Judge. (For example, 
%Powered by a public algorithm, it assesses data from the auxiliary database to detect potential privacy breaches. 

\textbf{Blockchain}: As mentioned earlier, our framework employs the 
blockchain to address the challenge of lie of the user. 
% Why the blockchain is a critical part of the Evidence Generation System has been explained in the previous 
The blockchain maintained by each aforementioned Web3 community 
records the relevant user behavior data (e.g., every service request 
sent to the platform,  
every response from the platform, every `stop using my data' request 
sent to the platform). 
In order to ensure user privacy, only the hash values 
of these user behavior data are stored in the blockchain. 
The original behavior data are separately kept by 
each individual user. 
%It safeguards the integrity of user requests by storing their hash values, preventing unauthorized access. 
The blockchain provides the Judge with the following 
{\em trustworthy property}: the hash values are immutable. 
Accordingly, when the Judge later receives any  
original behavior data from an user, the Judge can 
leverage the corresponding hash values (stored in the blockchain) 
and this property to verify that the user did not lie. 
Note that the hash values (and the corresponding 
original behavior data) are also major evidences 
the Evidence Generation System will later present to the Judge. 

Since the Auxiliary Database is also owned by the Web3 community, 
every update of this database must also be recorded in the 
blockchain. Otherwise, if the Auxiliary Database does not have 
transparency and immutability, the entire 
Evidence Generation System may become untrustworthy. 

\textbf{Judge}: 
%Serving as an impartial entity, the judge determines the validity of a user's request to stop the usage of their data based solely on the evidence presented.
After the Evidence Generation System sends all the major 
evidences to the Judge, the Judge will 
examine the collected evidences and decide if the accused is guilty. 

%-----------------
\noindent{\bf E-commerce illustration of our framework. }
E-commerce platforms provide users with merchant purchasing services. 
When our framework is employed to provide such users 
with better data right protection, the primary `stop using my data" 
concern is as follows: while the recommendation systems deployed 
on E-commerce platforms regularly use users' collective purchase 
histories (i.e., what is purchased and when) 
to make recommendations to individual users, 
each user has the right to ask any  
E-commerce platform to stop using his or her 
purchasing activities data when making any recommendations 
to any users. 

%To enhance comprehension without sacrificing generality, we consider a scenario in which users utilize the Amazon platform online to make purchases. 

In this scenario, the Server in our framework specifically 
refers to the E-commerce website.  
%utilizing the platform to buy items. Amazon incorporates a recommendation system that recommends new items to users, hence 
The Target System refers to the aforementioned 
recommendation system. 
The Database refers to the repository of purchase histories 
and the similarity map utilized by the recommendation system. 
% with the collaborative filtering algorithm being the algorithm employed by this target system. The sensitive information stored in the {\em database} comprises the purchase history of the users and the item-to-item similarity map which generated by the {\em target system}. 
Consider a group of users coming together to establish a Web3 community aimed at safeguarding their rights. When Alice (denoted as The Victim User in our framework), 
One of the community members, suspects that the E-commerce platform still 
uses her purchase history even after receiving her 
`stop using my data' request, she will ask the Evidence 
Generation System, a novel type of system proposed in this paper, 
to generate convincing evidences of data right violation.  

As we will shortly present in Section \ref{sec:case}, 
the specific Detector helping  
E-commerce users will utilize a community-owned  
similarity map, denoted as Auxiliary Database in our framework, 
to choose the items to purchase when probing the platform. 
Moreover, in order to generate the similarity map 
in a privacy-preserving manner, a standard MPC 
protocol will be employed by the group of users 
forming the Web3 community. 
%and contribute to the metadata using the collaborative filtering algorithm to generate the item-to-item similarity map, which is the metadata stored in the {\em auxiliary database}. 

%To prevent Alice from lying, the evidence generation system requests her to put the hash of her purchase history and request to the {\em blockchain}. After {\em detector} analyzing the metadata in the auxiliary database as well as the information Alice provided, the {\em evidence generation system} submits the generated evidence and the blockchain record to the {\em judge} for check. 

%In this example, we clarify the design and function of the overall system. Subsequent sections will offer detailed analyses of each component.

% As the data contains sensitive information, the user requests the system to stop using his or her purchase history. Our system allows users to form an MPC group with their family and friends. The MPC group suggests data based on the group's purchasing history, which is then sent to the detector. If a user suspects that their information is still being used, they submit their purchase history to the detector and provide the output of the MPC group to form an auxiliary database of tracked data. The detector then analyse these tracked data records into the blockchain and analyzes Amazon's use of the user's data based on the data submitted by the user. 

%In the following sections, we will analyze each component's structure and its functions one by one.

%-----------------------
\subsection{Multi-party Computation Group}
\label{section: MPC}

To address the data accessibility challenge 
%\ref{challenge: data} (disparity in accessible data), 
we propose the Web3 community that involves aggregating data from multiple users who share similar characteristics. However, this approach raises privacy concerns (e.g. the leakage of data between users), and to mitigate these concerns, we employ multiparty secure computation techniques to protect users' data. 

In our framework, we jointly compute the target algorithm (i.e., the algorithm used by the Target System) over users' private inputs while ensuring the confidentiality of this data. To deal with different target algorithms, our framework may employ different MPC techniques (e.g., homomorphic encryption, oblivious transfer, garbled circuits). 

Considering the {\em collaborative filtering algorithm} in a recommendation system 
example, which primarily utilizes multiply and add functions, homomorphic encryption emerges as a suitable MPC technique for conducting the necessary calculations. 
In this scenario, users encrypt their individual data using the CKKS fully homomorphic encryption algorithm (support both add and multiply functions) and collaboratively run the collaborative filtering algorithm on the encrypted data. 
While maintaining data confidentiality, we can obtain the same results 
as running the algorithm on cleartext data. 
% For the user group using our constructed evidence generation system, we choose to use homomorphic encryption to process their data and construct the system function $\mathbb{F}$ as a computation logic, in order to obtain an auxiliary dataset based on this user group. In this system, the evidence generation can only obtain information from the system output. For example, considering the target system as recommendation system, the output of the MPC group should be the similarity map obtained by the item-to-item collaborative filtering algorithm, which is currently the mainstream algorithm for recommendation systems.  

%--------------
\subsection{Evidence Generation System}

The evidence generation system forms the core of our framework. 
The system includes the auxiliary database, the detector, and the blockchain. 
% The auxiliary database is built using the outputs from the MPC Group and personal data of victim users, primarily designed to supply information to the detector. 
The auxiliary database is used to address the \textbf{data accessibility} challenge, which 
is built using the outputs from the MPC group.  
%and personal data of victim users, 
%primarily designed to supply information to the detector.

The detector algorithm is specifically designed for a particular 
target system to generate evidences. %Considering the  of target system, 
The {\em key insight} of Detector is as follows: 
Through modification of user inputs and monitoring the resultant 
outputs, we can ascertain whether the target system 
has leveraged the user's behavioral data. 
By utilizing the metadata stored in the auxiliary database, the detector can 
%gain insights into data relationships, 
predict the potential outputs for a specific input.  
Then, the predictions may be used by the Detector to 
%After altering the input provided by victim users, the detector 
assess the appropriateness of the observed outputs.

% Based on figure \ref{fig:sys}, the input of the evidence generation system includes 1) the output of the MPC group; 2) the information from the victim user. The output of the MPC group comprises metadata sourced from the Web3 community, stored in an {\em auxiliary database} to address data accessibility challenges. The information from the victim users includes the inputs from the users (e.g., the items victim users purchased) and the outputs from the target system (e.g., the items recommended). 

To address the \textbf{lie of user} challenge, our framework utilizes the blockchain. 
%In order to prevent users from feeding false data to the evidence generation system, it is essential to record users' service requests and the information they provide, while also ensuring the security and integrity of the data. 
As mentioned earlier, the blockchain is employed to tackle 
several types of dishonesty: (a) The user falsely asserted a violation of their data rights as a pretext to seek compensation. (b) The user provides false data to mislead the evidence generation system. (c) The user falsely claimed to have submitted a `stop using' request to the platform, but actually did not. 
Because the data (e.g., hash values) stored in the blockchain are immutable, 
all the data corruption (and deletion) attempts beyond dishonesty will fail.

Although the design of auxiliary databases and blockchains 
remains largely the same across various (types of) target systems, the 
design of the Detector may need to be target-system-specific. 
\section{Case Study: Recommendation Systems}
\label{sec:case}

Recommendation systems play a vital role in driving user engagement and business performance by delivering personalized recommendations based on user data analysis. These systems exploit user behavior and preferences to generate personalized suggestions. 
%emphasizing the importance of quality data in optimizing user experiences. 
We choose recommendation systems as our case study due to two reasons: 1) they are a most representative
type of 
% satisfies the two properties of a 
target systems in our framework; 2) they are widely employed 
by a variety of Internet-based platforms,  
%in our daily life, 
holding significant societal implications.

% Recommendation systems serve as pivotal tools in various domains due to their profound impact on user engagement and business performance. These systems are widely recognized for their ability to deliver personalized recommendations tailored to individual preferences, thereby facilitating informed decision-making and enhancing user satisfaction. Central to the functionality of recommendation systems is the utilization of user data, which serves as the foundation for generating accurate and relevant recommendations. By analyzing user behavior, preferences, and historical interactions, recommendation systems can discern patterns and correlations to predict user preferences effectively. Consequently, the success and effectiveness of recommendation systems hinge upon the availability and quality of user data, highlighting the pivotal role of data in driving personalized recommendations and optimizing user experiences.

%-------------------------
\subsection{Collaborative Filtering Algorithm}

Collaborative filtering is a widely-used algorithm in recommendation systems to provide users with personalized recommendations based on their historical preferences and behaviors \cite{sarwar2001item, linden2003amazon}. It works by analyzing the collective past interactions between users and items (movies, products, books, etc.) and then recommending new items to an user based on the similarities 
between the items previously-bought by the user and the items outside the user's purchase history. 
% the users' preferences and behaviors. 
The collaborative filtering algorithm has become an integral component of contemporary recommendation systems due to its widespread application across various industries. Its ability to analyze user preferences and behaviors has made it a go-to solution for businesses in e-commerce, online streaming, social media, and other domains. 

For a recommendation system with $M$ users and $N$ items, the rating of 
user $i$ on item $a$ is defined as $r_{i,a}$. 
(We assume that user $i$ will rate every item in his or her purchase history.) 
In accordance with the item-based collaborative filtering algorithm, the 
item vector is defined as $v_{a} = \{r_{0,a}, r_{1,a},...,r_{M,a}\}$.

The collaborative filtering algorithm contains three steps: 
1) Calculate the similarity between different items, among which $i \in I$ summations are over 
the users that have rated both the items $a$ and $b$: 

\centerline{$s_{a,b} = \frac{\sum_{i\in I}r_{a,i} r_{b,i}}{\sqrt{\sum_{i\in I}\!r_{a,i}^2} \sqrt{\sum_{i\in I}\!r_{b,i}^2}}$}

2) Let user $i$ be the target user. For each item $a$ not 
in the user's purchase history, calculate the ``how likely user $i$ will buy this item''
score for $a$ as follows. Here, 
%for each item in the purchase history of target user $i$, where 
the summations are over all other rated items $b \in H_i$; $s_{b,a}$ is the similarity between 
$b$ and $a$; $r_{i,b}$ is the rating of user $i$ on $b$.

\centerline{$score_{i,a} = \frac{\sum_{b \in H_i} s_{b,a}r_{i,b}}{\sum_{b \in H_i}|s_{i,b}|}$}\label{func:score}

3) Sort the scores calculated in the previous step 
in descending order and obtain the top $K$ highest 
scores as recommendations for user $i$. 

%The assembly of item 
It is clear that the similarity values between items 
serve as the basis for generating recommendations. 
Accordingly, the recommendation system will maintain a 
unique database called \textbf{similarity map}. 
%a core component of the recommendation system that inherently varies across databases. , the architecture of this similarity map is a cardinal concept in a recommendation system.

% The efficacy of a recommendation system is closely linked to the user inputs. Consequently, alterations to these inputs invariably affect the system's outputs. 
%Our objective is to establish the range within which the output can fluctuate, allowing us to identify any outputs that indicate the system's improper use of user information.

The interplay between the ``Stop Using My Data" right and the collaborative filtering algorithm 
is as follows: As soon as the recommendation system receives the ``Stop Using My Data" request 
from Alice, the similarity map should be revised in such a way that all the 
influence of Alice's purchase history on the similarity map should be removed. 
In addition, Alice's purchase history should no longer be used when 
making recommendations to Alice. 

% \textcolor{red}{can move these definitions to the 5.1 key insights parts}
% \begin{definition}[Distance] For the recommender system with similarity map $s$, the distance between item $a$ and $b$ is defined as: $d_{a,b}=1-s_{a,b}$\end{definition}

% The concept of `distance' refers to the disparity between different entities. Consider an example: a paper and a pen have lesser disparity than a paper and skincare products. This concept can inform the probability of particular recommendations, positing that, upon a user's purchase of paper, the system is more likely to suggest a pen over skincare products. However, recommendation systems seldom rely on the suggestion of a singular item. The system, therefore, necessitates gauging differences among a multitude of items. To satisfy this requirement, we introduce `cluster' as a conceptual framework to effectively evaluate relationships amidst these various items.

% \begin{definition}[$N$-Cluster] In a recommender system characterized by a similarity map, denoted as $s$, we classify the top-N $s_{a,i}$ items as the N-cluster for item $a$.
% \end{definition}

% By definition, a cluster represents a collection of objects in proximity to a given item, with the measurement of proximity determined by the concept of distance. Based on our assumptions, the platform uses similarity map forms the $N$-cluster of the purchase history, then randomly chooses $K$ items as recommendations to the user.

\begin{figure}[t]
    \centering  \includegraphics[width=8cm]{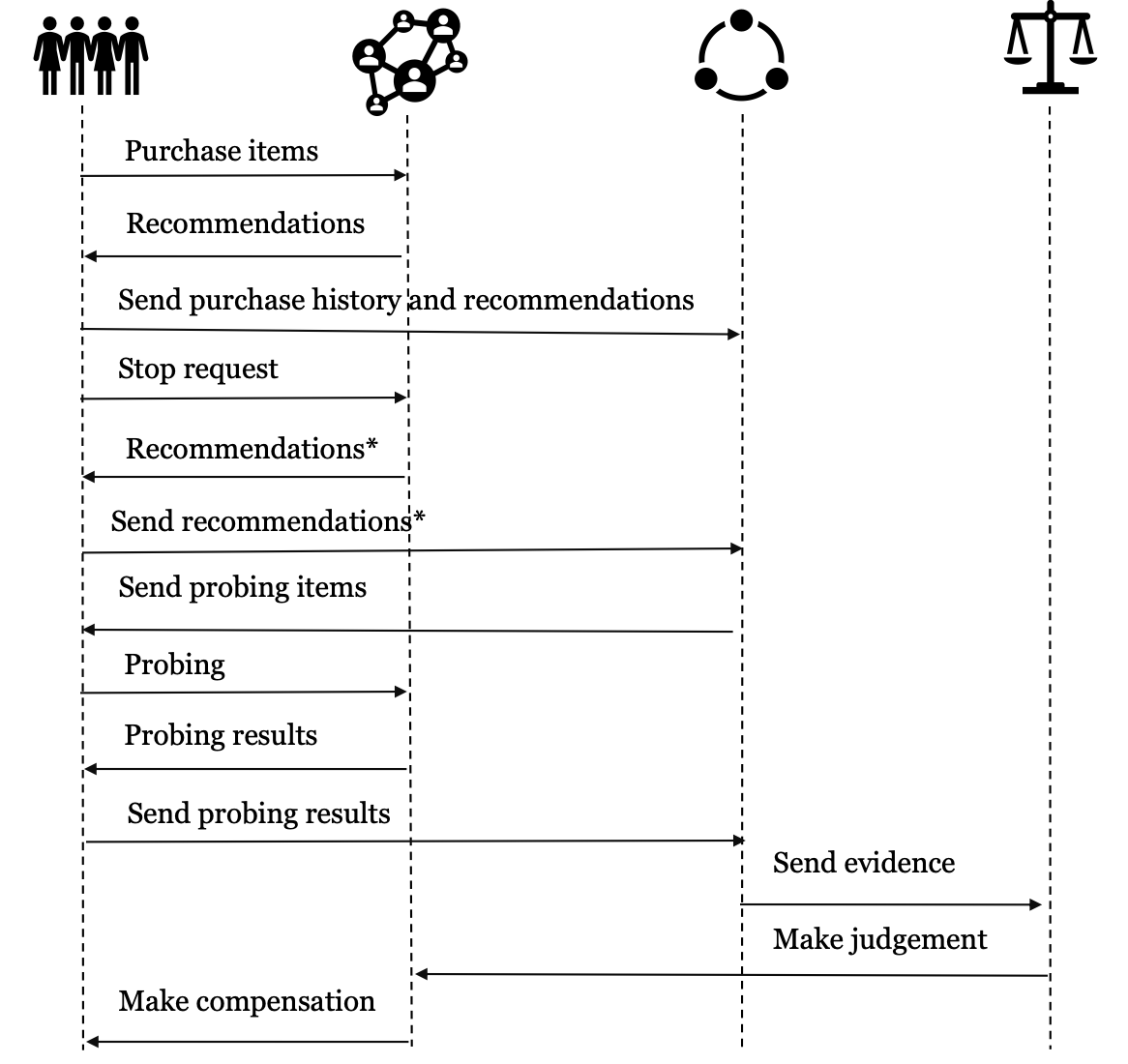}
    \caption{The evidence generation dataflow.}
    \label{fig:dataflow}
    \vspace{-0.2in}
\end{figure}

%-----------------------------
\subsection{Evidence Generation System Overview}

Figure \ref{fig:dataflow} illustrates how evidences are 
generated through the interactions between 
four main components: user, platform (recommendation system), 
evidence generation system, and a judge. 
Initially, a user, Alice, purchases items on the platform, which subsequently analyzes her purchase history to generate the recommendations. 
As a member of the Web3 community, Alice shares 
both her purchase history and the past recommendations with 
the evidence generation system. 
After Alice submits a request asking the platform to 
discontinue the use of her data, the data right violator 
platform continues to generate new recommendations using her purchase history. 
As soon as Alice senses data right violation, 
% Questioning the platform’s adherence to her request, 
she shares the new recommendations with the evidence generation system. 
The system uses a novel probing algorithm, which we will present shortly 
in next section, to gather the corresponding evidences. 
As soon as adequate evidences are generated, they will be 
sent to the Judge. 
If the evidences suggest that the platform has indeed improperly used the 
user’s purchase history, the Judge 
may mandate that the platform provide compensation.

%------------------
\section{Two-Phase Evidence Generation Protocol}

%--------------------------
\subsection{Key Insights}

% The performance of the recommendation system is highly dependent on user input. As a result, any alteration made to the input will be reflected in the system's output. Our objective is to establish the range within which the output can fluctuate, allowing us to identify any outputs that indicate the system's improper use of user information.

% \textcolor{red}{this section should be more specific}

% Our system design allows the judge to receive information from four different sources. The user provides his purchase history(which should be protected), along with the output of the platform(recommendations) to the detector. The Web3 community provides the auxiliary similarity map, which, along with the user's data, form the auxiliary database. With the auxiliary dataset, 

% \textbf{Main Challenge} 
% Despite the information provided by the victim user and the MPC group, it remains challenging to ascertain whether the system has ceased utilizing the user's data.  

\begin{figure}[htbp]
    \centering
    \includegraphics[width=3.6in]{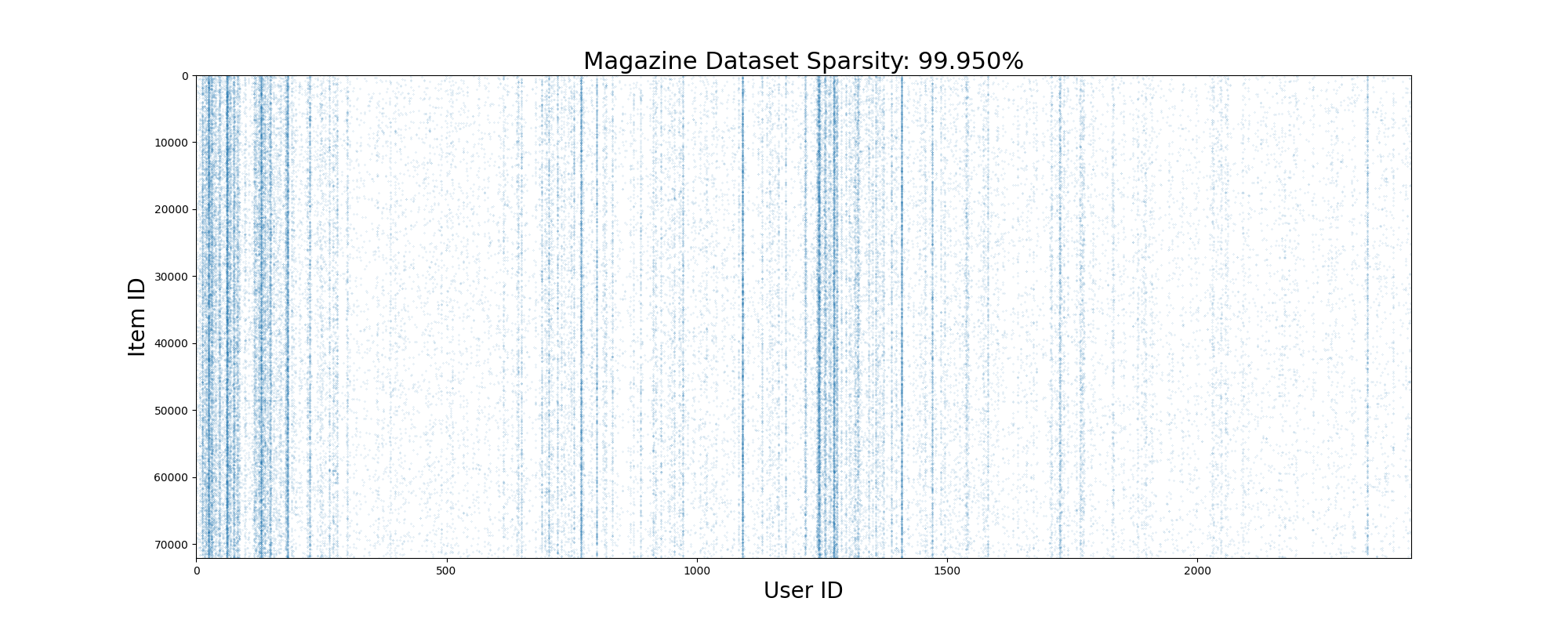}
    \vspace{-0.3in}
    \caption{Amazon Magazine dataset Sparsity.}
    \label{fig:data1}
    \vspace{-0.1in}
\end{figure}
% \vspace{-0.3in}

\begin{figure}[htbp]
    \centering
    \includegraphics[width=3.6in]{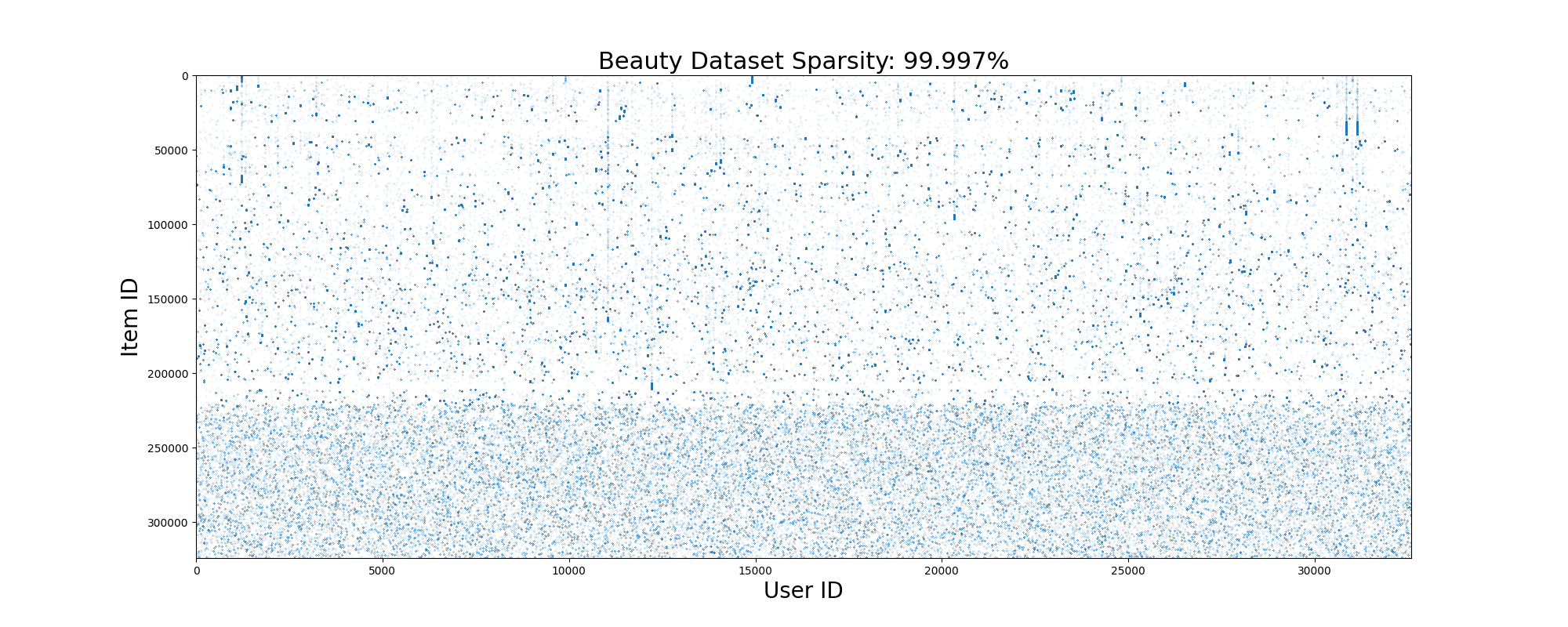}
    \vspace{-0.3in}
    % \vspace{-2mm}
    \caption{Amazon Beauty dataset Sparsity.}
    \label{fig:data2}
    \vspace{-1mm}
\end{figure}

\textbf{Key observation.} 
Figure \ref{fig:data1} and Figure \ref{fig:data2},  
obtained through the analysis of the real-world 
Amazon Magazine dataset and the Amazon 
Beauty dataset, 
%platform Amazon and platform Netflix, 
respectively, provides a key observation, that is, 
real-world similarity maps have {\em high sparsity}. 
The high sparsity property is manifested by the observation 
that a considerable portion of items are 
  only ever bought by a limited number of users. 

%Hence, the inclusion of an item's record in the auxiliary dataset strongly suggests that it encompasses all records pertaining to that particular item. 

%-------------------------
\textbf{Key insight.}  
The key data sparsity property of real-world similarity maps indicates that 
although the auxiliary similarity map (i.e., auxiliary database) maintained 
by the evidence generation system is calculated 
based on the purchase histories of 
a limited number of users in the Web3 community, 
the map may already hold all the similarity relationships 
for many items included in the map. 
%The limited number of item records in the real database is due to its characteristic sparsity, resulting in the auxiliary database containing all records for certain items. 
Consequently, the recommendations derived from the auxiliary similarity map 
for an item are very likely to be accurate, making them suitable for 
probing the platform.  

%Based on the aforementioned ideas, we primarily consider the following two core questions: Firstly, \textbf{how to choose the target item}, how can we effectively identify and select target items in the auxiliary database that holds the same purchase records in the real database? Secondly, \textbf{the evaluation of system outputs}, how can we determine whether the output is based on the user's purchase history?

%Based on our assumptions, we are unable to access information from the real database, thereby hindering our ability to establish precise mathematical upper and lower bounds for the two problems. To better address this issue, we propose a two-phase evidence generation algorithm in this section, which analyzes and provides approximate solutions to these two problems. 

%-----------------------------
\subsection{The Evidence Generation Protocol}

Our evidence generation protocol comprises two distinct stages. %to address specific issues. 
%The study showcases the ability of our evidence generation system to protect data rights. 
The two stages answer the following two research questions, respectively. 
% focused on answering the following two research questions: 
%Our focus in this section is to tackle the inquiries outlined in Figure \ref{fig:problem}.
% \newtcolorbox{mybox}{colframe = blue!75!black}
%\begin{tcolorbox}[title=\textbf{Two-phase Problems}, colback=red!5,colframe=red!75!black]

{\bf (Q1)} The real similarity map is only partially known to the evidence 
generation system. How to choose the probing items 
in such a way that the impact of limited knowledge (about the 
similarity map) is minimized? 

{\bf (Q2)} How to reason whether the gathered evidences are adequate? 
%platform has stopped using the data? 
%\end{tcolorbox}

We use the following example %in Figure \ref{fig:prob} 
to illustrate our protocol:  
\begin{tcolorbox}[title = {Running Example}, colback=Salmon!10, colframe=Salmon!80!Black]
Alice purchases some items on a platform. 
Her purchase history includes stationery like pencil, eraser, pen and books 
like Harry Potter. Based on her purchase history, the platform 
recommends a few other stationery and books to her. 
To protect her data right, she sends a ``Stop Using My Data'' 
request to the platform. 
\end{tcolorbox}

%----------------------
\subsection{Choose the Probing Items}

Based on the definition of similarity between items 
and the aforementioned key observation, we make the following two 
basic assumptions: 
% We answer the first question under two foundational assumptions:
\begin{assumption}
The smaller the similarity (values) between items, the larger the 
differences (between items) are. 
\label{as:sim}
\end{assumption} 

\begin{assumption}
Without knowing the similarity map maintained by the 
recommendation system, utilization of the auxiliary similarity map  
enables the evidence generation system to 
approximately evaluate the similarities between items.
\label{as:meta}
\end{assumption}

% \subsubsection{The introduction of cluster} 
We found that appropriate probing items can be chosen based 
on several criteria. However, in order to 
avoid ambiguity in specifying these criteria, 
two notions must be firstly defined. 

%To better understand the assumption \ref{as:sim}, we first introduce the definition of cluster:
\begin{definition}[per-item-cluster] 
    For each item $x$ bought by a member of the Web3 community, 
    we define the corresponding per-item-cluster based on the 
    following operations. 
    
\textbf{Initialization.} At the inception of the Web3 community, some 
users already possessed their purchase histories. 
%This data, utilized in conjunction with Multi-party Computation (MPC), facilitated the creation of an 
These histories will be used to calculate the initial auxiliary similarity map. 
Then, for each item $x$ in the auxiliary similarity map, 
we let the initial per-item-cluster of $x$ include every other item (in the map) 
that has non-zero similarity between itself and $x$. 
% from the auxiliary map to establish the initial.

%A cluster refers to a set of items grouped 
%together based on similarity, which leads to 
%recommendations for the objects within the collection.
%Subsequently, we will delineate the development of the cluster and elucidate its evolutionary process.
%\textbf{Initialization} At the inception of the Web3 community, some users already possessed their purchase history. This data, utilized in conjunction with Multi-party Computation (MPC), facilitated the creation of an auxiliary similarity map. For each item, denoted as (p), we identified and selected all items demonstrating non-zero similarity from the auxiliary map to establish the initial (p)-cluster.

\textbf{The cluster update operations. } 
We may need to update certain per-item-clusters when an    
item $a$ is purchased.  
The update operations will handle two different cases. 
%involving two distinct types of updates. 
(Case 1) When a new user makes item $a$ her first purchase, 
we immediately update the per-item-cluster of $a$ to include 
all newly recommended items. 
(Note that if the per-item-cluster does not exist, 
it will be created with $a$ and all recommended items.) 
(Case 2) If $a$ is not the buyer's first purchase, 
for each item $p$ previously bought by the user, 
we calculate $sim(a, p)$ and update the corresponding value 
in the auxiliary similarity map. 
% If the similarity is non-zero, we update the auxiliary similarity map and incorporate this item into the per-item-cluster of $a$.   
Then, we will merge the user's purchase history 
into the per-item-cluster of $a$. 
Next, we will update the per-item-clusters of each 
item $p$ previously bought by the user. Note that some of these 
clusters may not contain $a$. 
Finally, we will merge the newly recommended items 
into the per-item-cluster of $a$. 

%---------------
\textbf{The cluster growing operations.}   
When an item $a$ is purchased, we not only need to perform 
the above update operations, but only need to merge 
the newly recommended items into every 
per-item-cluster that contains $a$. 
Regarding why this merge operation is needed, note that 
%Given that the Web3 community encompasses only a subset of users, 
since the auxiliary similarity map differ from the actual similarity map, 
items may exhibit zero similarity (with $a$) in the auxiliary map 
while demonstrating non-zero similarity in the real map due to incomplete 
set of purchases used in calculating the auxiliary map.
%(the real cluster and the auxiliary cluster will be different in this scenario). 
In such cases, the real and auxiliary per-item-clusters will differ. 
This merge operation is performed to 
let the evidence generation system achieve a better approximation 
of the black-box target system. 
%If item $p$ appears in any recommendation for an item $x$ which is not in $p$-cluster, those items should be added to the cluster.
\end{definition} 

\begin{figure*}[t]
    \centering  \includegraphics[width=15cm]{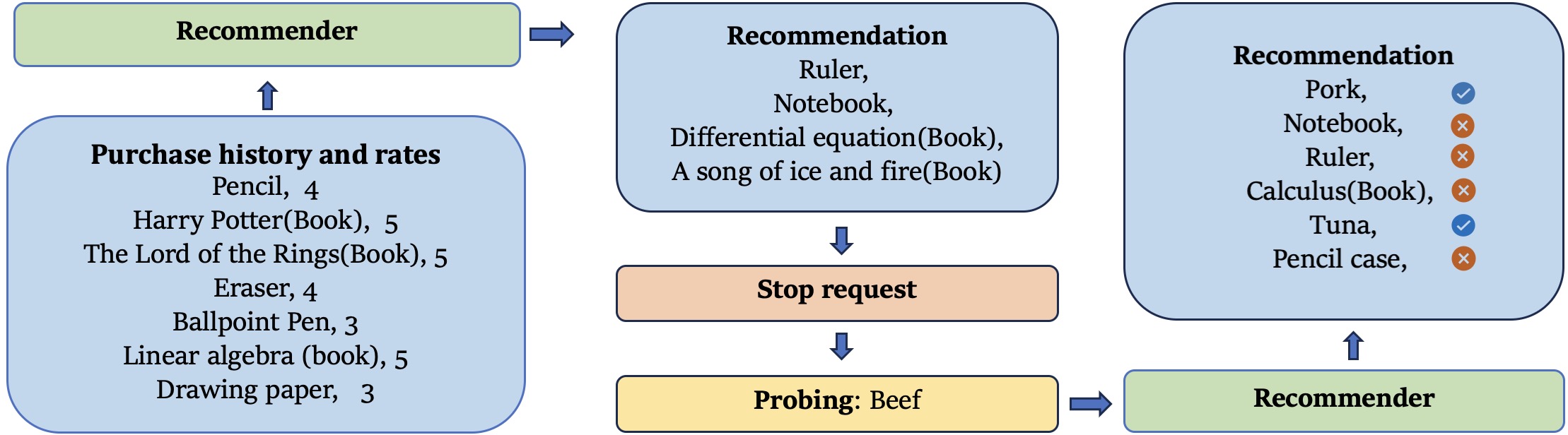}
    \caption{The probing process.}
    \label{fig:prob}
    \vspace{-4mm}
\end{figure*}

%\begin{remark}
\begin{definition}[Purchase History Cluster]
    We amalgamate all the  
    per-item-clusters for each item in the purchase history of user $i$ 
    into an union termed the purchase history cluster of user $i$. 
\end{definition}
%\end{remark} 

%------------------------------
{\bf Classify the users in the Web3 community into two groups.} 
%Since we only provide help for the users in the Web3 community, for the other users we unfortunately do not help them. 
We categorize the users in the Web3 community into two groups, based on the congruence 
between recommendations derived from the auxiliary similarity map 
and those from the actual recommendations. 
We classify the users whose received recommendations for all 
purchased items are already in the per-item-clusters 
calculated based on the auxiliary similarity map 
%coincide with auxiliary clusters 
as belonging to group one, while those 
experiencing discrepancies fall into second group. 
%\subsubsection{How we choose the probing items}

Now we propose the following 
{\bf criteria for choosing the probing items}: 
%We choose the probing items based on four constraints: 
(C1) A chosen item should be the 
first item purchased by a non-victim user in 
the Web3 community. 
(C2) The observed platform recommendations for a chosen item 
should entirely fall within the item’s auxiliary cluster which is 
calculated based on the auxiliary similarity map.  
(C3) The per-item-cluster for a chosen item should contain no item 
from any growing operations. 
(C4) The similarity between a chosen item and 
the items in the victim user’s purchase history should be zero.  

For each victim user in the Web3 community, the evidence generation system can obtain her entire purchase history 
and all the recommendations from the platform. 
For example, assume Alice’s purchase history is [a, b]. We know the 
recommendations for history [a] and the recommendations for history [a, b]. 
We cannot get the recommendations for history [b] as item $a$ is purchased before $b$.  
In this situation, 
criterion C1 helps to ensure that the 
recommendation generated based on the auxiliary similarity map is 
the recommendation provided by the platform. 
Criteria C2 and C3 guarantee the 
discrepancy between the auxiliary and the actual clusters is 
minimal or even nonexistent. 
% Considering D1, 
Note that it is highly probable, if not certain, that the recommendations 
for the probing item will fall within the auxiliary cluster. 
Based on the aforementioned basic assumptions, 
criterion C4 helps us get a clear difference between 
the recommendations for the probing item alone and the recommendations 
for Alice's purchase history plus the probing item.

Based on the collaborative filtering algorithm, score is calculated 
by multiplying similarity and ratings, 
where ratings are directly assigned by users. 
To make this difference even bigger, 
%between the recommendations for the purchase history plus the probing item 
%and [probing item] larger, 
we let Alice assign rating value 1.0, which is the highest rating, to the probing item.  
Note that different rating values will not influence the recommendations 
for the probing item, but the recommendations for the purchase history plus the probing item 
will be influenced. The smaller the assigned rating value is, the larger the history vs. no history difference. Using rating value 1.0, this difference would be minimized. 

%------------------------------------
\subsection{Using the probing items to generate adequate evidences}

The probing process is illustrated in Figure \ref{fig:prob}. 
We answer the second research question Q2 based on the following assumption:

\begin{assumption}
    The recommendations from the platform form a cluster (and we assume we know the size of it), but we only know part of the cluster based on the output of the platform. We denote such clusters as an R cluster. For example, the R cluster size is 10, but each time the recommendation system will randomly choose 7 items as the recommendations. \label{as:cluster} 
\end{assumption}

\begin{tcolorbox}[colback=SeaGreen!10!CornflowerBlue!10,colframe=RoyalPurple!55!Aquamarine!100!,title = {Answer for Q1}]
%Building upon the previous approach, 
We first employ criterion C1 to identify all the candidate items. 
Second, we employ criterion C4 to filter off the first set of unqualified items. 
Third, we employ criterion C3 to filter off the second set of unqualified items.  
Finally, we employ criterion C2 to filter off the last set of unqualified items.  
%We firstly determine Alice's membership in Group one. Subsequently, we assess the per-item-clusters associated with her complete purchase history, encompassing items such as pencils, Harry Potter, eraser, etc. Subsequently, we select an item from our target cluster that meets all four constraints, ultimately identifying `beef' as the probing item. 
\end{tcolorbox}

    Although we assume we know the cluster size, the cluster size is sometimes difficult to infer. 
    %a deviation from reality. 
    For example, in real-world platforms like Amazon, recommended products vary with each page refresh or following a re-login. To handle the scenarios in which the evidence generation system 
    cannot know the cluster size, we assume that the Judge is willing to obtain such metadata from the recommendation system.  
    
        %We assume that all recommended products are part of a single cluster, from which the platform randomly selects a predetermined number of products for recommendation during each instance. Determining the size of this cluster proves to be a challenging task.

Considering criterion C4, which stipulates that (the union of) the auxiliary clusters of the items in the victim’s purchase history and the R cluster of the probing item have no intersection, and given the above assumption, which allows us to determine the R cluster size of the probing item, we can draw the following conclusion: 

\vspace*{1mm}
\begin{lemma}
    \label{lemma_1}
Let's denote the (set of) disclosed recommendations for a probing item $p$ as set B.  
Let's denote the per-item-cluster of $p$ as set A. 
If the cardinality of $A-B$ 
exceeds the total number of undisclosed recommendations, it can be 
concluded that the platform continues utilizing the victim user's purchase history. 
    %If the disclosed recommendations for a probing item $p$ include items not from the auxiliary clusters, and the total number of {\em such} items exceeds the total number of undisclosed recommendations, it can be concluded that the platform continues utilizing the victim user's purchase history. 
\end{lemma} 

{\bf Proof.} For this lemma, 
our ``proof by contradiction'' is as follows: Let's assume that the platform 
had stopped utilizing the victim user's purchase history. 
If some recommended items don’t belong to the per-item-cluster of $p$, this 
must be due to 
incomplete knowledge held by the evidence generation system, 
and therefore such items should be added  
into the per-item-cluster of $p$. 
However, after these items are added,  
the size of the per-item-cluster would 
be larger than the known size, which 
contradicts Assumption \ref{as:cluster}. 

Since the observation mentioned in Lemma 6.1 is not guaranteed to be observed
through only one probing item, there is 
usually a need to leverage two or more probing items. 
In our evaluation experiments (See Section VII), we will measure 
on average how many probing items are needed. 

%\begin{lemma}
%    \label{lemma:2}
Besides Lemma 6.1, the following observation may make 
the evidences more convincing: 
If the recommendations for the probing item include items 
from the victim’s purchase history, it is very likely that the 
platform still uses the victim user's purchase history.
%\end{lemma}

According to criteria C2 and C4, there is no intersection between the 
known portions of the cluster for the probing item and the purchase history.  
Given that the similarity between the probing item and items in the purchase history is zero, the existence of an item that maintains non-zero similarity with 
both the probing item and purchase history, yet remains undetected by the 
Web3 community, is extremely rare. However, under assumption \ref{as:cluster}, since our 
knowledge of the per-item-cluster of the probing item and 
the purchase history cluster remains incomplete, we cannot 
conclude that the possibility (of not using the victim user's history)
is zero. 
%Nonetheless, it remains possible.

\begin{remark}
\label{remark:1}
For a user in group two, the similarity between the user's  
purchase history and the probing items may not be zero. 
So, it is harder for us to get enough evidence as 
the difference may not be big enough. 
\end{remark}

\begin{tcolorbox}[colback=SeaGreen!10!CornflowerBlue!10,colframe=RoyalPurple!55!Aquamarine!100!,title = {Answer for Q2}]
%When being probed, the platform's output includes several items that do not belong to the purchase history cluster, such as notebooks, rulers, and books. Since the total cluster size is known, and the quantity of these items surpasses the unknown size, 
Whenever the observation described in Lemma 6.1 is observed, 
we can deduce that the platform  
continues utilizing Alice's purchase history.   
\end{tcolorbox}

\subsection{Blockchain Design}
In this section, we introduce our current design of blockchain. In blockchain, each user has a public key and a private key. The public key is used to receive transactions, while the private key is used to sign transactions. The public key is visible to anyone, but the private key is kept confidential and known only to the user. Therefore, while transactions can be traced, the true identity of the transaction users remains confidential.

The whole process of the web3 community member contains three steps: 1) Transaction initiation: After a user purchases an item, they create a transaction containing payment information, details of the purchased item, and the transaction time. They use their private key to sign a digital signature, which is then appended to the end of the currency, creating a transaction record. 2) Transaction broadcast: This transaction is broadcasted to the blockchain network. Nodes in the network verify the validity of this account. 3) Smart contract execution: If the transaction is verified as valid, smart contracts on the blockchain update the data, and the transaction is added to a new block on the blockchain. This block is appended to the end of the blockchain, creating a permanent and immutable record. 

The smart contract code is as following, we define one structure Transaction, which records the purchase of user and two functions, which are used for recording purchase and getting transaction.

\begin{lstlisting}
pragma solidity ^0.5.16;
contract PurchaseRecord {
    struct Transaction {
        address buyer;
        string itemID;
        uint256 rate;
        uint256 transactionTime;
    }
    Transaction[] public transactions;
    function recordPurchase(string memory itemDetails, uint256 paymentAmount) public {
        Transaction memory newTransaction = Transaction({
            buyer: msg.sender,
            item: itemID,
            rate: rate,
            transactionTime: now
        }); transactions.push(newTransaction);
    }
    function getTransaction(uint256 index) public view returns (address, string memory, uint256, uint256) {
        Transaction memory transaction = transactions[index];
        return (transaction.buyer, transaction.itemID, transaction.rate, transaction.transactionTime);
    }
}
\end{lstlisting}

The whole process can be easily applied to existing blockchain like Ethereum, by compiling the source code of the smart contract into bytecode then transmitting the bytecode to public nodes. 

\subsection{Multi-party Computation}

In this section, we explain how we apply muti-party computation into the web3 community. We implement item-based collaborative filtering algorithm in homomorphic encryption to ensure the privacy of user purchases and ratings. We apply Cheon-Kim-Kim-Song(CKKS) as the homographic methods. CKKS is a homomorphic encryption scheme known for its ability to perform computations on floating-point operations with low noise growth. However, as CKKS cannot perform square root operations, therefore, we will divide the above section into two parts for the calculation: 

\textbf{Initialization Phase}
Each user $i$ has a private key $sk_i$ and a public key $pk_i$ generated using a homomorphic encryption scheme. The rating of user $i$ on item $a$, denoted as $r_{i,a}$, is encrypted as $Enc_{pk_i}(r_{i,a})$ and sent to the server. 

Also, as CKKS can not calculate square root operation now. 

\textbf{Similarity Computation Phase}
The server computes the similarity between different items in an encrypted form. For items $a$ and $b$, the similarity score $s_{a,b}$ is computed based on ratings that users have given to both items, with the following modification for secure computation:

1) The encrypted similarity score between item $a$ and $b$ is calculated by the server using the homomorphic properties of the encryption scheme: 
\\
\centerline{ $ Enc(s_{a,b}) = \frac{\bigoplus_{i\in I} Enc_{pk_i}(r_{i,a}) \otimes Enc_{pk_i}(r_{i,b})}{\sqrt{\bigoplus_{i\in I}\!(Enc_{pk_i}(r_{i,a})^2)} \cdot \sqrt{\bigoplus_{i\in I}\!(Enc_{pk_i}(r_{i,b})^2)}} $},  

where $\oplus$ denotes homomorphic addition, and $\otimes$ denotes homomorphic multiplication.
    
2) The server then sends the encrypted similarity score $Enc(s_{a,b})$ to a secure entity capable of decrypting the information or to the users themselves for secure decryption.

\textbf{Decryption and Normalization Phase}
Finally, the similarities are decrypted and normalized to be used for generating recommendations.

\begin{enumerate}
    \item The secure entity or users (whoever is responsible for decryption) decrypt $Enc(s_{a,b})$ using their corresponding private keys to obtain the similarity score $s_{a,b}$.
    \item The decrypted similarity scores can then be used for generating recommendations while ensuring the privacy of $r_{i,a}$ is maintained throughout the computation process.
\end{enumerate}

\section{Evaluation}

%This section analyzes the evidence generation system by conducting experiments on the approximated components of the algorithm using real-world datasets. The results of these experiments support our initial hypotheses and strengthen the credibility of our system design. Considering that blockchain technology can be directly implemented on a public chain, we have chosen not to include its experimental results in this section.

Our evaluation experiments will focus on answering the following 
questions: (Q1) Are the probing items selected by our 
evidence generation system indeed the ones intended? 
Since the real similarity map is only partially known to 
our evidence generation system, the probing items might be 
mistakenly selected. (Q2) For each user in Group One, 
how many rounds of probing are needed for the 
evidence generation system to obtain the platform output that   
adheres to Lemma 6.1? Note that as soon as such an output 
is obtained, the Judge can be convinced via the lemma.
(Q3) How large is the portion of group one users 
in a representative Web3 community? Note that group one users
are the targets of protection of our evidence generation system. 

Besides answering the three questions, we also evaluated
the time consumed to compute the auxiliary similarity maps through MPC. 
Nevertheless, 
since platform probing is triggered by  
violations of ``Stop Using My Data'' and since such violations
are usually infrequent, measuring the end-to-end processing time of 
our evidence generation system is not a major concern of 
the Web3 communities. Accordingly, we did not measure 
the processing time of the blockchain in our system.

\subsection{Dataset}
We use two real-world datasets, namely Amazon Magazine Dataset and Amazon Beauty Dataset. The Magazine Dataset contains 89,688 (purchasing) records with 72,098 items from 2,428 users and the Beauty dataset contains 371,344 records with 324,038 items from 32,586 users. We illustrate the sparsity of these two datasets by presenting their heat-maps in Figure \ref{fig:data1} and Figure \ref{fig:data2}, respectively, where the x-axis represents the users and the y-axis represents the items. The data indicate that both datasets exhibit extremely high levels of sparsity. Specifically, the Amazon Magazine dataset demonstrates a sparsity of 99.997\%, whereas the other dataset shows a sparsity of 99.95\%.

% \subsection{Similar clusters or not?}
% Leveraging our preliminary assumptions on cluster analysis, we employ an auxiliary database for computation, recognizing the potential for approximation. We then evaluate the discrepancies between the real dataset and the auxiliary dataset in relation to target items.

% \subsection{Blockchain}
% We compare our blockchain design with the state-of-art Ethurem. 

\subsection{Are the adopted probing items indeed the ones intended?}

\begin{table*}[]
\centering
\scriptsize
\caption{The comparison between the clusters obtained through auxiliary similarity map vs. real similarity map of Amazon Magazine dataset.}
\vspace{-1mm}
\label{tb_cluster1}
\begin{tabular}{@{}c|cc|cc|cc|cc@{}}
\toprule
\multirow{2}{*}{\textbf{Size of Web3 Community}} & \multicolumn{2}{c|}{\textbf{2\% User}} & \multicolumn{2}{c|}{\textbf{5\% User}} & \multicolumn{2}{c|}{\textbf{10\% User}} & \multicolumn{2}{c}{\textbf{20\% User}} \\ \cmidrule(lr){2-9}
 & Auxiliary & Real & Auxiliary & Real & Auxiliary & Real & Auxiliary & Real \\ \midrule
\textbf{Target Item Number} & \multicolumn{2}{c|}{39} & \multicolumn{2}{c|}{92} & \multicolumn{2}{c|}{192} & \multicolumn{2}{c}{406}  \\ \midrule
\textbf{\begin{tabular}[c]{@{}c@{}}Target Item with\\  Same Clusters\end{tabular}} & \multicolumn{2}{c|}{38 (97.4\%)} & \multicolumn{2}{c|}{91 (98.9\%)} & \multicolumn{2}{c|}{187 (97.4\%)} & \multicolumn{2}{c}{392 (96.6\%)}\\ \midrule
% \textbf{The Minimum Size} & 1 & 1 & 1 & 1 & 1 & 1 & 1 & 1 \\
% \textbf{The Maximum Size} & 50 & 1835 & 71 & 165 & 165 & 1157 & 1340 & 3249 \\
\textbf{The Average Size} & 10.4 & 56.2 & 13.0 & 14.7 & 15.1 & 22.4 & 24.0 & 46.5 \\
\textbf{\begin{tabular}[c]{@{}c@{}}The Average Size of\\  Same Clusters\end{tabular}} & \multicolumn{2}{c|}{9.3} &  \multicolumn{2}{c|}{13.0} & \multicolumn{2}{c|}{14.8} & \multicolumn{2}{c}{18.8} \\
\bottomrule
\end{tabular}
\vspace{-4mm}
\end{table*}

\begin{table*}[]
\centering
\scriptsize
\caption{The comparison between the clusters obtained through auxiliary similarity map vs. real similarity map of Amazon Beauty dataset.}
\vspace{-1mm}
\label{tb_cluster2}
\begin{tabular}{@{}c|cc|cc|cc|cc@{}}
\toprule
\multirow{2}{*}{\textbf{Size of Web3 Community}} & \multicolumn{2}{c|}{\textbf{2\% User}} & \multicolumn{2}{c|}{\textbf{5\% User}} & \multicolumn{2}{c|}{\textbf{10\% User}} & \multicolumn{2}{c}{\textbf{20\% User}} \\ \cmidrule(lr){2-9}
 & Auxiliary & Real & Auxiliary & Real & Auxiliary & Real & Auxiliary & Real \\ \midrule
\textbf{Target Item Number} & \multicolumn{2}{c|}{526} & \multicolumn{2}{c|}{1323} & \multicolumn{2}{c|}{2676} & \multicolumn{2}{c}{5466}  \\ \midrule
\textbf{\begin{tabular}[c]{@{}c@{}}Target Item with\\  Same Clusters\end{tabular}} & \multicolumn{2}{c|}{505 (96.0\%)} & \multicolumn{2}{c|}{1284 (97.1\%)} & \multicolumn{2}{c|}{2611 (97.6\%)} & \multicolumn{2}{c}{5353 (97.9\%)}\\ \midrule
% \textbf{The Minimum Size} & 1 & 1 & 1 & 1 & 1 & 1 & 1 & 1 \\
% \textbf{The Maximum Size} & 49 & 2626 & 359 & 735 & 399 & 2793 & 8347 & 8723 \\
\textbf{The Average Size} & 6.1 & 14.2 & 6.7 & 10.6 & 6.8 & 13.5 & 11.4 & 16.0 \\
\textbf{\begin{tabular}[c]{@{}c@{}}The Average Size of\\  Same Clusters\end{tabular}} & \multicolumn{2}{c|}{5.7} &  \multicolumn{2}{c|}{6.4} & \multicolumn{2}{c|}{6.5} & \multicolumn{2}{c}{9.4} \\
\bottomrule
\end{tabular}
\vspace{-4mm}
\end{table*}

\begin{table}[]
\centering
\scriptsize
\caption{The percentage of probing items whose clusters are totally different from the victim's purchase history cluster.}
\label{tab_2}
\begin{tabular}{@{}c|c|c|c|c@{}}
\toprule
\textbf{\begin{tabular}[c]{@{}c@{}}Size of Web3 \\ Community\end{tabular}} & \textbf{2\% User} & \textbf{5\% User} & \textbf{10\% User} & \textbf{20\% User}\\   \midrule
\textbf{Amazon Magazine} & 100.0\% & 100.0\% & 100.0\% & 100.0\% \\
\textbf{Amazon Beauty} & 100.0\% & 100.0\% & 100.0\% & 100.0\% \\
\bottomrule
\end{tabular}
\vspace{-4mm}
\end{table}

The selection of probing items is a crucial step, greatly impacting 
the evidence generation process. 
We denote every item which meets all of the four 
criteria (i.e., C1, C2, C3, and C4)
presented in Section VI a {\em target item}. 
In Section VI, we hypothesize that the recommendations derived from 
the auxiliary similarity map for target items are
very likely to be accurate, making them suitable for probing the platform.
In this section, we will test this hypothesis 
through the following comparison: 
\begin{itemize}
\item For each of the two real-world datasets we use, we build 
four representative Web3 communities, holding 2\% of users, 
5\% of users, 10\% of users, and 20\% of users, respectively. 
\item We replay the purchasing activities 
recorded in the real-world dataset and run the 
Collaborative Filtering Algorithm to construct the 
real similarity map and use it to generate recommendations. 
\item When the purchasing activities are being replayed, 
we let the evidence generation system of each Web3 community to simultaneously construct  
each community's auxiliary similarity maps and use the auxiliary maps 
to maintain all the per-item-clusters for each community.  
%identify all the target items 
Then we use the four criteria presented in Section VI to 
identify the target items for each community. 
\item Assume that an oracle community knows 
the real similarity map (at any point of time), we let the 
oracle community use the real map to 
maintain the per-item clusters for the items purchased 
by each Web3 community. 
%the real-world dataset. 
Then we use the per-item clusters to identify 
the target items for each Web3 community.
\item Now, we can compare the target items identified using 
the auxiliary similarity maps and the target items identified using the real
similarity map. 
Note that the aforementioned hypothesis would be True if the comparison 
only reveals minimum differences. 
\end{itemize}

The main evaluation results are as follows. 
First, in terms of which target items are identified, the results 
are shown in the top lines of Table \ref{tb_cluster1} and Table  
\ref{tb_cluster2}, respectively. 
The results show that the same set of target items 
are identified with or without using the real similarity map. 
The results also show that the percentage of target items remains stable as 
the Web3 community expands.

%whereas the percentage of all items in the real database increases with the growth of the auxiliary database. 

%Given that the real similarity map is kept confidential  
% operates as a black box 
%and our assessments rely on data from an auxiliary , there are primarily two significant information gaps:

%1). During the selection of the target items, we assume that if all recommendation items (based on real dataset) also belong to the cluster of the target item (based on auxiliary dataset), then the information of that item in the auxiliary dataset is the same as in the real dataset. However, this assumption can introduce some inaccuracy.

%2). When selecting probing items for a specific user from the target items, it is assumed that if the cluster for the probing item contains no item from any growing operations and the similarity between the probing item and the items in the victim user’s purchase history is zero, then there is a substantial difference in their outputs. However, due to differences between the real and auxiliary databases, this assumption may introduce certain gaps. 

%To address the previously mentioned information gaps, a series of experiments were undertaken using the Amazon dataset. From this dataset, a random sample of users, consisting of $[2\%, 5\%, 10\%, 20\%]$, was selected to form the Web3 community. The items identified as targets by our algorithm were then evaluated for information discrepancies in both the auxiliary and real datasets. 

Second, we assess the differences between the pair of per-item clusters for each target item. 
In Table \ref{tb_cluster1} and \ref{tb_cluster2}, we can see the percentage of the target items holding the same pair of clusters is always higher than 96\%. 
The results also show that the size of clusters tends to be small for target items. 
% Larger clusters are associated with a higher probability of exhibiting differences. 
Third, we measured the percentage of the probing items whose per-item clusters are totally different from the victim's purchase history cluster. As shown in the two tables, 
the percentages are 100\% in all of the four Web3 communities. 
%Both results prove the gap in the probing items part is very small. 
%Besides, in Table \ref{tab_2}, to evaluate the gap 2, 

% \begin{table}[]
% \label{tab:1}
% \centering
% \caption{The accuracy of the target items.}
% \begin{tabular}{@{}l|cccc@{}} 
% \toprule
% Percentage of users & 2\%  & 5\%  & 10\% & 20\% \\\hline
% Amazon Magazine     & 97.4 & 98.9 & 96.8 & 96.3 \\\hline
% Amazon Beauty       & 100.0 & 99.9 &  100.0  & 100.0  \\ \bottomrule

% \end{tabular} 
% \end{table}

% \begin{table}[]
% \centering
% \caption{The percentage of
% probing items holds totally different clusters with purchase
% history.}
% \label{tab_2}
% \begin{tabular}{@{}l|cccc@{}} \hline
% \textbf{Size of Web3 Community} & 2\%  & 5\%  & 10\% & 20\% \\\hline
% Amazon Magazine     & 100.0 & 100.0 & 100.0 & 100.0 \\\hline
% Amazon Beauty       & 100.0 & 100.0 & 100.0 & 100.0    \\\hline
% \end{tabular} 
% \end{table}

% \begin{figure}[t]
%     \centering  \includegraphics[width=2.8in]{fig/cluster.png}
%     \caption{The accuracy of target items.}
%     \label{fig:sys}
% \end{figure}
% \subsection{Do probing item clusters and purchase history clusters have empty intersection?}

\subsection{How many users are in group one?}
% \begin{table}[htbp]
% \centering
% \caption{The percentage of target users.}
% \label{tab:3}
% \begin{tabular}{l|cccc} \hline
% \textbf{Size of Web3 Community} & 2\%  & 5\%  & 10\% & 20\% \\ \hline
% Amazon Magazine     & 39.6 & 33.1 & 34.7& 33.4  \\ \hline
% Amazon Beauty       & 64.4 & 67.4 & 68.4 & 70.4 \\\hline
% \end{tabular} 
% \end{table}

\begin{table}[]
\centering
\scriptsize
\caption{The percentage of group one users %and the analysis of purchase history 
of Amazon Magazine dataset.
}
\label{tb_user_1}
\begin{tabular}{@{}c|c|c|c|c@{}}
\toprule
\textbf{\begin{tabular}[c]{@{}c@{}}Size of Web3 \\ Community\end{tabular}} & \textbf{2\% User} & \textbf{5\% User} & \textbf{10\% User} & \textbf{20\% User}\\   \midrule
\textbf{Group One User} & 39.6\% & 33.1\% & 34.7\%& 33.4\% \\ \midrule
\textbf{Minimum Length} & 1 & 1 & 1 & 1  \\
\textbf{Maximum Length} & 12 & 12 & 15 & 15 \\
\textbf{Average Length} & 3.4 & 3.0 & 2.8 & 2.8 \\
\bottomrule
\end{tabular}
\end{table}

\begin{table}[]
\centering
\scriptsize
\caption{The percentage of group one users %and the analysis of purchase history 
of Amazon Beauty dataset. }
\vspace{-1mm}
\label{tb_user_2}
\begin{tabular}{@{}c|c|c|c|c@{}}
\toprule
\textbf{\begin{tabular}[c]{@{}c@{}}Size of Web3 \\ Community\end{tabular}} & \textbf{2\% User} & \textbf{5\% User} & \textbf{10\% User} & \textbf{20\% User}\\   \midrule
\textbf{Group One User} & 64.4\% & 67.4\% & 68.4\% & 70.4\% \\ \midrule
\textbf{Minimum Length} & 1 & 1 & 1 & 1  \\
\textbf{Maximum Length} & 38 & 38 & 38 & 48 \\
\textbf{Average Length} & 2.6 & 2.6 & 2.6 & 2.7 \\
\bottomrule
\end{tabular}
\vspace{-4mm}
\end{table}

% the number of target users: \\
% 20\%: 162/485 ~~~~
% 10\%: 84/242 ~~~~
% 5\%: 40/121 ~~~~
% 2\%: 19/48 ~~~~

%Based on the protocol, users are divided into two categories, with the first category comprising the target users. 
As mentioned in Section VI, users are divided into two 
categories: group one users vs. group two users. 
Since group one users are the targets of protection of our evidence generation system, 
we measure whether representative Web3 communities have numerous group one users or 
only a few. Tables \ref{tb_user_1} and \ref{tb_user_2} show the proportion of 
group one users within each Web3 community. 
The results show that a substantial 
proportion (ranging from 33.1\% to 39.6\%) of users are indeed group one   
users in each Web3 community. 
In addition, the results show that 
despite the growth of the Web3 community from 2\% to 20\% of users, 
the proportion stays stable. 
%ranging from 33.1\% to 39.6\%. This demonstrates that 
%the protocol's effectiveness regardless of its size. The analysis also revealed that the purchase histories of target uses are typically brief, averaging 2 to 3 items. This is because longer histories tend to have a higher likelihood of including items with incomplete records in the auxiliary database. 

Tables \ref{tb_user_1} and \ref{tb_user_2}
also show the length statistics of the purchase histories of group one users. 
Since longer histories tend to have a higher likelihood of including items 
with incomplete information in the auxiliary similarity map, it is not 
very surprising that the histories are usually short. 

\subsection{How many group one users can be successfully protected?}

\begin{table}[]
\centering
\scriptsize
\caption{The success rate of platform probing %process and the percentage of outputs from purchase history. 
}
\vspace{-1mm}
\label{tb_success_rate}
\begin{tabular}{@{}c|c|c|c|c@{}}
\toprule
\textbf{\begin{tabular}[c]{@{}c@{}}Size of Web3 \\ Community\end{tabular}} & \textbf{2\% User} & \textbf{5\% User} & \textbf{10\% User} & \textbf{20\% User}\\   \midrule
\textbf{\begin{tabular}[c]{@{}c@{}}Successful Rate \\ Magazine\end{tabular}} & 100.0\% & 100.0\% & 99.5\%& 99.7\% \\ 
\textbf{\begin{tabular}[c]{@{}c@{}}Successful Rate \\ Beauty\end{tabular}} & 100.0\% & 99.9\% & 99.3\%& 99.5\% \\ 
%\textbf{\begin{tabular}[c]{@{}c@{}}Items-Magazine \end{tabular}} & 41.6\% & 37.8\% & 44.6\%& 63.5\%  \\
%\textbf{Items-Beauty} & 19.2\% & 19.4\% & 18.9\% & 19.1\% \\
\bottomrule
\end{tabular}
\vspace{-4mm}
\end{table}

We first define the one-round probing success rate 
as follows: if the next probing item we use is randomly (i.e., uniform 
distribution) selected from the set of target items, the success rate
is defined as the probability that the selected 
target item will trigger the platform to generate 
the output that adheres to Lemma 6.1. 

In order to measure this probability, 
for each user in Group One, we treated the user as the victim and 
evaluated all the target items for the user. 
Let's assume there are in total $N$ target items. 
Let's assume the number of the target items 
which trigger the platform to generate 
the output that adheres to Lemma 6.1 is $M$. 
Then the probability is $M/N$. 
The results in Table \ref{tb_success_rate} show that 
the one-round probing success rates are 
exceedingly high, approaching 100\%. 
In case the first round of probing does not succeed, 
the second round of probing will actually enjoy 
a higher succeed rate since the two rounds of probing events
are independent. 
%that meet the constraints of the probing protocol. 
%If the output adheres to Lemma \ref{lemma_1}, the probing process is deemed successful. We observed that the success rates are exceedingly high, approaching 100\%.

%Additionally, we analyzed the proportion of items in the outputs that belong to auxiliary purchase history clusters, which also exhibits a high percentage. This observation is consistent with remark \ref{remark:1}. 

%-----------------------
\subsection{The Multi-party Computation}

We apply the CKKS homomorphic encryption from Microsoft SEAL to compute 
the auxiliary similarity map while 
protecting the privacy of the behavioral data held by each user locally.  
%information of both users and items. 
We compare the times consumed to compute the similarity map  
with and without MPC. For the Amazon Magazine dataset, we selected 2\% of users to represent our Web3 community. The computation time for the similarity map is 8 minutes and 42 seconds without CKKS, and 56 minutes and 44 seconds with CKKS. 
Regarding why we selected 2\% of users, we notice that in real-word scenarios, 
instead of including all users in a very large Web3 community, such a Web3 community is often 
built by users with similar purchasing behavior. Therefore, such a Web3 community 
is usually small.

\section{Discussion}
%Data protection is an ongoing and pressing issue that has endured neglect for an extended period. In this article, we conducted experiments to test the use of the web3 community for safeguarding the data rights of community members. We implemented different protection measures for various users, and the experimental results demonstrate the effectiveness of our algorithm even for data points with missing information. 

%---------------------
\subsection{Discussion of Grey-box Assumption}

During the construction of our evidence generation system, we make
%simplifications, including 
the following assumption: We assume that the Internet-based platform 
is a grey-box, meaning that the data is largely unknown but the platform 
functions are known. 
% In the following chapters, we will delve into a comprehensive discussion of the aforementioned two aspects.

%Within the parameters of our research framework, it is postulated that the platform's database remains an unknown entity, whereas the functions of the platform are acknowledged. 
%This assumption serves to streamline the complexities inherent in the system and enhances the potential for theoretical verification. 

This assumption intends to serve two purposes: (a) It enables researchers to 
capture a primary technical challenge in solving the evidence generation 
problem. That is, the evidence generation system neither has access to 
the behavioral data stored in the platform, nor knows which portion of the data 
is used when responding to a particular user service request. Since 
(a portion of) the behavioral data is used to compute each response 
generated by the platform, without the ability 
to utilize the data, it is very difficult to determine whether 
a response is violating the ``Stop Using My Data'' right or not. 
(b) The assumption captures an important pre-condition 
for a Web3 community to gain the needed evidence generation ability. 
That is, even if the Web3 community enables a substantial portion 
of the behavioral data stored in the platform to become 
accessible to an evidence generation system 
with strong privacy and integrity protection, 
the evidence generation system still needs to know 
how responses are computed by the platform. 
Otherwise, it seems that the platform can always deny 
right violation accusations by the following argument: 
``Because the evidences are generated based on a  
different way to compute the response, the evidences 
are not very relevant to the accusation.'' 

%The central aim of the grey-box assumption is to attain verifiability. Though it may marginally enhance the success rate, we contend that as long as the dependency between inputs and outputs remains unchanging, we can ensure the efficacy of our algorithm.

Regarding whether this assumption is reflecting the 
real-world, let's take recommendation systems as an example. 
On one hand, we found that the algorithm employed by 
many real-world recommendation systems is indeed based on  
collaborative filtering. 
On the other hand, we found that platform functions are sometimes
not a white-box. 
%sometimes a black-box. 
For example, a recommendation system leverages user purchase history and profiles to generate recommendations, employing algorithms that transcend basic collaborative filtering and incorporate the aid of artificial intelligence. 
In such cases, the assumption indicates a gap between 
the evidence generation system proposed in this paper 
and real-world platforms. 
Nevertheless, we found that this gap can be bridged in 
a realistic way. First, the Judge has the authority 
to obtain the platform functions. Second, 
even if the platform functions stay unknown to the 
Web3 community, the evidence generation system 
can let the platform functions running on the Judge side 
to access the data maintained by the Web3 community. 
In this way, the evidence generation system does not need to be 
redesigned. 

%\textbf{Our primary hypothesis, postulating that `the system output is solely dependent on and determined by user input',  continues to hold validity.} Therefore, the two steps of the two-phase evidence generation, namely determine the target item and use the target item for probing will remain effective as it relies on the system output for evaluation.

% \subsection{Discussion of the honest system assumption}
% In this study, we make the assumption that the system operates with integrity, implying that the output of the system is solely determined by its algorithm and the inputs provided by the user. No alterations are made to manipulate or avoid scrutiny. In practical situations, in order to avoid monitoring, recommendation systems are prone to employing a variety of deceitful tactics, such as selectively including or excluding users' historical data in their recommendations. Moreover, it is not limited to recommendation systems alone; other target systems may also exhibit different forms of evasive conduct. 

% We argue that \textbf{the aforementioned issue can be effectively represented as the Byzantine Generals' Problem across diverse systems}. Consequently, when encountering multiple times of probings, adequate evidence can suffice to impose penalties on the target system. The extent of fault tolerance rates can be derived based on the specific forms of dishonest behavior exhibited by different systems.

\subsection{Discussion of Cross-user Scenarios}

In this paper, our analysis focuses on scenarios in which the platform utilizes user data to generate recommendations tailored to individual users. 
However, we note that the system has the potential to utilize user data for recommending to other users as well. 
%In the current study, we focus solely on scenarios where the system employs user data to make recommendations to the same user. However, the system is also capable of utilizing one user's data to generate recommendations for a different user
We denote such a scenario as ``cross-user'' scenarios. 
Even within a single Web3 community, detecting 
violations of ``Stop Using My Data'' in cross-user 
scenarios remains challenging. 
This difficulty arises since such violations involve 
data dependencies between two users' purchase histories.  
%disrupt the internal correlation of the data, and the platform's database is essentially a black box to us. 
Since the data stored in the platform is largely unknown 
to the Web3 community, the data dependencies 
sometimes cannot be fully analyzed. 
%There are also problems of information asymmetry in other target systems. 
%Due to the mismatch of data volume, 
For example, let's assume that Alice and Bob are 
members of the Web3 community, but George is not. 
Without knowing the purchase activities of George, 
we may be unable to determine whether Alice's purchase 
history {\em indirectly} depends upon Bob's history through 
George's purchase activities.  
%similar records to that of the user. Only when all users belong to the web3 community can we determine whether cross-user data right violations exist.

% \subsection{Relationship between our finding and differential privacy}
% Differential privacy is a technical method that protects individual privacy by adding noise to data queries, ensuring that the presence or absence of any individual data point does not substantially affect the query results. 

%Additionally, our experimental findings indicate that in practical systems, the effectiveness of differential privacy is likely to diminish when applied to specific users and items. Thus, we have conducted a comprehensive analysis and discussion regarding this issue.

%Differential privacy has played an important role in protecting specific individual information. However, our findings indicate that in real-world systems, differential privacy may not be applicable. For example, adding noise can impact the output of recommendation systems. Additionally, our results complement differential privacy by suggesting that it is no longer applicable when the database has sparsity. 

%Due to the in applicability of differential privacy, we can acquire information through system probing and determine system data by querying the results. Subsequently, an evidence generation system can be constructed based on the system data to assess whether user data has been utilized.

\section{Conclusion and Future Work}
\label{sec:con}

To the best of our knowledge, 
this paper introduces the first evidence generation system for protecting the 
``Stop Using My Data'' right. 
%We initially address the challenges in supervising the `right to be forgotten' and 
We first present a technical definition of the 
``Stop Using My Data'' right in the context of Internet-based platforms. 
%stop using my data' right. 
Then, based on blockchain technology and MPC technology, we 
propose the first evidence generation framework for potential victims 
receiving services from Internet-based platforms. 
%The system analyzes the sensitivity of different data and provides different levels of protection. 
The framework has a set of unique characteristics. 
Towards validating the proposed evidence generation framework, 
we conduct a case study on recommendation systems 
with systematic evaluation experiments using 
two real-world datasets (i.e., the Amazon Magazine dataset 
and the Amazon Beauty dataset). 

%as a study case, analyzed the influence of auxiliary dataset, the difference of data sensitivity, and the impact of the dataset on the system efficiency, and confirmed the effectiveness of the system design. 

In future research, it is imperative to broaden the application of the 
proposed evidence generation framework. In our future work, 
we aim to design evidence generation systems not only for 
e-commerce platforms, but other kinds of platforms such as Gboard, a platform 
providing various typing services.  
%applicable platforms. 
Furthermore, the current evidence generation system lacks 
the capability to safeguard one user's data from being leveraged in 
generating the platform's responses for other users. 
In our future work, we aim to extend our evidence generation 
system to handle ``cross-user'' scenarios. 

%such as the inclusion of a user's data in recommendations made to their friends within a recommendation system. Currently, we have not yet conducted testing of our system on a real, industry-leading platform like Amazon. The exploration of these concerns warrants thorough analysis and discussion in future investigations. 

\bibliographystyle{IEEEtran}
\bibliography{sample.bib}

\begin{thebibliography}{10}
\providecommand{\url}[1]{#1}
\csname url@samestyle\endcsname
\providecommand{\newblock}{\relax}
\providecommand{\bibinfo}[2]{#2}
\providecommand{\BIBentrySTDinterwordspacing}{\spaceskip=0pt\relax}
\providecommand{\BIBentryALTinterwordstretchfactor}{4}
\providecommand{\BIBentryALTinterwordspacing}{\spaceskip=\fontdimen2\font plus
\BIBentryALTinterwordstretchfactor\fontdimen3\font minus
  \fontdimen4\font\relax}
\providecommand{\BIBforeignlanguage}[2]{{%
\expandafter\ifx\csname l@#1\endcsname\relax
\typeout{** WARNING: IEEEtran.bst: No hyphenation pattern has been}%
\typeout{** loaded for the language `#1'. Using the pattern for}%
\typeout{** the default language instead.}%
\else
\language=\csname l@#1\endcsname
\fi
#2}}
\providecommand{\BIBdecl}{\relax}
\BIBdecl

\bibitem{sha2020survey}
K.~Sha, T.~A. Yang, W.~Wei, and S.~Davari, ``A survey of edge computing-based
  designs for iot security,'' \emph{Digital Communications and Networks},
  vol.~6, no.~2, pp. 195--202, 2020.

\bibitem{mayrhofer2021android}
R.~Mayrhofer, J.~V. Stoep, C.~Brubaker, and N.~Kralevich, ``The android
  platform security model,'' \emph{ACM Transactions on Privacy and Security
  (TOPS)}, vol.~24, no.~3, pp. 1--35, 2021.

\bibitem{Termly2024}
\BIBentryALTinterwordspacing
M.~Komnenic, ``64 alarming data privacy statistics businesses must see in
  2024,'' 2024. [Online]. Available:
  \url{https://termly.io/resources/articles/data-privacy-statistics/}
\BIBentrySTDinterwordspacing

\bibitem{rocha2020functionality}
T.~Rocha, E.~Souto, and K.~El-Khatib, ``Functionality-based mobile application
  recommendation system with security and privacy awareness,'' \emph{Computers
  \& Security}, vol.~97, p. 101972, 2020.

\bibitem{jung2020secure}
J.~Jung, J.~Cho, and B.~Lee, ``A secure platform for iot devices based on arm
  platform security architecture,'' in \emph{2020 14th International Conference
  on Ubiquitous Information Management and Communication (IMCOM)}.\hskip 1em
  plus 0.5em minus 0.4em\relax IEEE, 2020, pp. 1--4.

\bibitem{li2022design}
C.~Li, H.~Niu, M.~Shabaz, and K.~Kajal, ``Design and implementation of
  intelligent monitoring system for platform security gate based on wireless
  communication technology using ml,'' \emph{International Journal of System
  Assurance Engineering and Management}, pp. 1--7, 2022.

\bibitem{zaeem2020effect}
R.~N. Zaeem and K.~S. Barber, ``The effect of the gdpr on privacy policies:
  Recent progress and future promise,'' \emph{ACM Transactions on Management
  Information Systems (TMIS)}, vol.~12, no.~1, pp. 1--20, 2020.

\bibitem{voigt2017eu}
P.~Voigt and A.~Von~dem Bussche, ``The eu general data protection regulation
  (gdpr),'' \emph{A Practical Guide, 1st Ed., Cham: Springer International
  Publishing}, vol.~10, no. 3152676, pp. 10--5555, 2017.

\bibitem{bygrave2019minding}
L.~A. Bygrave, ``Minding the machine v2. 0: the eu general data protection
  regulation and automated decision making,'' \emph{Algorithmic Regulation
  (Oxford University Press 2019, Forthcoming, University of Oslo Faculty of Law
  Research Paper No. 2019-01}, 2019.

\bibitem{lecuyer2019certified}
M.~Lecuyer, V.~Atlidakis, R.~Geambasu, D.~Hsu, and S.~Jana, ``Certified
  robustness to adversarial examples with differential privacy,'' in \emph{2019
  IEEE Symposium on Security and Privacy (SP)}.\hskip 1em plus 0.5em minus
  0.4em\relax IEEE, 2019, pp. 656--672.

\bibitem{dong2022gaussian}
J.~Dong, A.~Roth, and W.~J. Su, ``Gaussian differential privacy,''
  \emph{Journal of the Royal Statistical Society Series B: Statistical
  Methodology}, vol.~84, no.~1, pp. 3--37, 2022.

\bibitem{alabdulatif2020towards}
A.~Alabdulatif, I.~Khalil, and X.~Yi, ``Towards secure big data analytic for
  cloud-enabled applications with fully homomorphic encryption,'' \emph{Journal
  of Parallel and Distributed Computing}, vol. 137, pp. 192--204, 2020.

\bibitem{das2021bim}
M.~Das, X.~Tao, and J.~C. Cheng, ``Bim security: A critical review and
  recommendations using encryption strategy and blockchain,'' \emph{Automation
  in construction}, vol. 126, p. 103682, 2021.

\bibitem{wang2022srr}
H.~Wang, H.~Hong, L.~Xiong, Z.~Qin, and Y.~Hong, ``L-srr: Local differential
  privacy for location-based services with staircase randomized response,'' in
  \emph{Proceedings of the 2022 ACM SIGSAC Conference on Computer and
  Communications Security}, 2022, pp. 2809--2823.

\bibitem{wang2023shuffle}
S.~Wang, X.~Luo, Y.~Qian, Y.~Zhu, K.~Chen, Q.~Chen, B.~Xin, and W.~Yang,
  ``Shuffle differential private data aggregation for random population,''
  \emph{IEEE Transactions on Parallel and Distributed Systems}, 2023.

\bibitem{politou2018forgetting}
E.~Politou, E.~Alepis, and C.~Patsakis, ``Forgetting personal data and revoking
  consent under the gdpr: Challenges and proposed solutions,'' \emph{Journal of
  cybersecurity}, vol.~4, no.~1, p. tyy001, 2018.

\bibitem{peloquin2020disruptive}
D.~Peloquin, M.~DiMaio, B.~Bierer, and M.~Barnes, ``Disruptive and avoidable:
  Gdpr challenges to secondary research uses of data,'' \emph{European Journal
  of Human Genetics}, vol.~28, no.~6, pp. 697--705, 2020.

\bibitem{huynh2023blockchain}
T.~Huynh-The, T.~R. Gadekallu, W.~Wang, G.~Yenduri, P.~Ranaweera, Q.-V. Pham,
  D.~B. da~Costa, and M.~Liyanage, ``Blockchain for the metaverse: A review,''
  \emph{Future Generation Computer Systems}, 2023.

\bibitem{bhutta2021survey}
M.~N.~M. Bhutta, A.~A. Khwaja, A.~Nadeem, H.~F. Ahmad, M.~K. Khan, M.~A. Hanif,
  H.~Song, M.~Alshamari, and Y.~Cao, ``A survey on blockchain technology:
  Evolution, architecture and security,'' \emph{Ieee Access}, vol.~9, pp.
  61\,048--61\,073, 2021.

\bibitem{li2020survey}
X.~Li, P.~Jiang, T.~Chen, X.~Luo, and Q.~Wen, ``A survey on the security of
  blockchain systems,'' \emph{Future generation computer systems}, vol. 107,
  pp. 841--853, 2020.

\bibitem{choi2022blockchain}
T.-M. Choi and T.~Siqin, ``Blockchain in logistics and production from
  blockchain 1.0 to blockchain 5.0: An intra-inter-organizational framework,''
  \emph{Transportation Research Part E: Logistics and Transportation Review},
  vol. 160, p. 102653, 2022.

\bibitem{damgaard2006scalable}
I.~Damg{\aa}rd and Y.~Ishai, ``Scalable secure multiparty computation,'' in
  \emph{Advances in Cryptology-CRYPTO 2006: 26th Annual International
  Cryptology Conference, Santa Barbara, California, USA, August 20-24, 2006.
  Proceedings 26}.\hskip 1em plus 0.5em minus 0.4em\relax Springer, 2006, pp.
  501--520.

\bibitem{alexopoulos2017mcmix}
N.~Alexopoulos, A.~Kiayias, R.~Talviste, and T.~Zacharias, ``Mcmix: Anonymous
  messaging via secure multiparty computation.'' in \emph{USENIX security
  symposium}, 2017, pp. 1217--1234.

\bibitem{yao1982protocols}
A.~C. Yao, ``Protocols for secure computations,'' in \emph{23rd annual
  symposium on foundations of computer science (sfcs 1982)}.\hskip 1em plus
  0.5em minus 0.4em\relax IEEE, 1982, pp. 160--164.

\bibitem{zhao2019secure}
C.~Zhao, S.~Zhao, M.~Zhao, Z.~Chen, C.-Z. Gao, H.~Li, and Y.-a. Tan, ``Secure
  multi-party computation: theory, practice and applications,''
  \emph{Information Sciences}, vol. 476, pp. 357--372, 2019.

\bibitem{fan2023altruistic}
S.~Fan, T.~Min, X.~Wu, and W.~Cai, ``Altruistic and profit-oriented: Making
  sense of roles in web3 community from airdrop perspective,'' in
  \emph{Proceedings of the 2023 CHI Conference on Human Factors in Computing
  Systems}, 2023, pp. 1--16.

\bibitem{nabben2023web3}
K.~Nabben, ``Web3 as ‘self-infrastructuring’: The challenge is how,''
  \emph{Big Data \& Society}, vol.~10, no.~1, p. 20539517231159002, 2023.

\bibitem{farahani2021convergence}
B.~Farahani, F.~Firouzi, and M.~Luecking, ``The convergence of iot and
  distributed ledger technologies (dlt): Opportunities, challenges, and
  solutions,'' \emph{Journal of Network and Computer Applications}, vol. 177,
  p. 102936, 2021.

\bibitem{zhu2019applications}
Q.~Zhu, S.~W. Loke, R.~Trujillo-Rasua, F.~Jiang, and Y.~Xiang, ``Applications
  of distributed ledger technologies to the internet of things: A survey,''
  \emph{ACM computing surveys (CSUR)}, vol.~52, no.~6, pp. 1--34, 2019.

\bibitem{zou2019smart}
W.~Zou, D.~Lo, P.~S. Kochhar, X.-B.~D. Le, X.~Xia, Y.~Feng, Z.~Chen, and B.~Xu,
  ``Smart contract development: Challenges and opportunities,'' \emph{IEEE
  Transactions on Software Engineering}, vol.~47, no.~10, pp. 2084--2106, 2019.

\bibitem{sarwar2001item}
B.~Sarwar, G.~Karypis, J.~Konstan, and J.~Riedl, ``Item-based collaborative
  filtering recommendation algorithms,'' in \emph{Proceedings of the 10th
  international conference on World Wide Web}, 2001, pp. 285--295.

\bibitem{linden2003amazon}
G.~Linden, B.~Smith, and J.~York, ``Amazon. com recommendations: Item-to-item
  collaborative filtering,'' \emph{IEEE Internet computing}, vol.~7, no.~1, pp.
  76--80, 2003.

\end{thebibliography}

\end{document}